\documentclass[twocolumn]{vautart}  
  
\usepackage{amsmath}
\usepackage{bbold}
\usepackage{amssymb}
\usepackage{graphicx}    
\usepackage{enumerate}
\usepackage{comment,color}
\usepackage{accents}
\usepackage{ulem}
\newtheorem{cor1}{\bf Corollary}
\newtheorem{lem1}{\bf Lemma}

\newtheorem{assum1}{\bf Assumption}
\newtheorem{rem1}{\bf Remark}

\usepackage{chngcntr}
\usepackage{apptools} 
\AtAppendix{\counterwithin{lem1}{section}}

\newcommand*{\QEDB}{\hfill\ensuremath{\square}}

\newcommand{\R}{\mathbb{R}}
\newcommand{\N}{\mathbb{N}}

\newcommand{\cE}{\mathcal{E}}

\newcommand{\cI}{\mathcal{I}}

\newcommand{\cG}{\mathcal{G}}

\newcommand{\cN}{\mathcal{N}}
\newcommand{\KL}{\mathcal{KL}}
\newcommand{\Kinf}{\mathcal{K}_\infty}

\newcommand{\eps}{\varepsilon}

\newcommand{\cM}{\mathcal{M}}

\newcommand{\FF}{\mathbf{F}}

\newcommand{\RR}{\mathbf{R}}
\newcommand{\ZZ}{\mathbf{Z}}
\newcommand{\WW}{\mathbf{W}}
\newcommand{\uu}{\mathbf{u}}
\newcommand{\xx}{\mathbf{x}}
\newcommand{\ee}{\mathbf{e}}
\newcommand{\yy}{\mathbf{y}}
\newcommand{\zz}{\mathbf{z}}

\newcommand{\dd}{\mathbf{d}}

\newcommand{\sss}{\mathbf{s}}
\newcommand{\ww}{\mathbf{w}}

\newcommand{\setdef}[2]{\left\{\ #1\ \left|\ \vphantom{#1} #2\ \right.\right\}}

\renewcommand{\subset}{\subseteq}

\definecolor{antiquefuchsia}{rgb}{0.57, 0.36, 0.51}
\definecolor{chromeyellow}{rgb}{1.0, 0.65, 0.0}
\definecolor{darkgreen}{rgb}{0,0.6,0}

\begin{document}

\begin{frontmatter}

\title{Edge-wise funnel output synchronization of heterogeneous agents with relative degree one\thanksref{footnoteinfo}}
\thanks[footnoteinfo]{This work was supported by the National Research Foundation of Korea(NRF) grant funded by the Korea government(Ministry of Science and ICT) (No. NRF-2017R1E1A1A03070342).
This work was done while Jin Gyu Lee was with Seoul National University.
}

\author[JGLee]{Jin Gyu Lee}\ead{jin-gyu.lee@inria.fr},
\author[TBerger]{Thomas Berger}\ead{thomas.berger@math.upb.de},
\author[STrenn]{Stephan Trenn}\ead{s.trenn@rug.nl},
\author[HShim]{Hyungbo Shim}\ead{hshim@snu.ac.kr}

\address[JGLee]{Inria, University of Lille, CNRS, UMR 9189 - CRIStAL, F-59000 Lille, France}
\address[TBerger]{Institut f{\"u}r Mathematik, Universit{\"a}t Paderborn, Warburger Stra{\ss}e~100, 33098~Paderborn, Germany}
\address[STrenn]{Department of Mathematics, University of Groningen, Netherlands}  
\address[HShim]{ASRI, Department of Electrical and Computer Engineering, Seoul National University, Korea}
          

\begin{abstract}
When a group of heterogeneous node dynamics are diffusively coupled with a high coupling gain, the group exhibits a collective emergent behavior which is governed by a simple algebraic average of the node dynamics called the blended dynamics.
This finding has been utilized for designing heterogeneous multi-agent systems by building the desired blended dynamics first and then splitting it into the node dynamics. However, to compute the magnitude of the coupling gain, each agent needs to know global information such as the number of participating nodes, the graph structure, and so on, which prevents a fully decentralized design of the node dynamics in conjunction with the coupling laws. 
To resolve this issue, the idea of funnel control, which is a method for adaptive gain selection, 
can be exploited for a node-wise coupling, but the price to pay is that the collective emergent behavior is no longer governed by a simple average of the node dynamics. 
Our analysis reveals that this drawback can be avoided by an edge-wise design premise, which is the idea that we present in this paper. After all, we gain benefits such as a fully decentralized design without global information, collective emergent behavior being governed by the blended dynamics, and the plug-and-play operation based on edge-wise handshaking between two nodes.
\end{abstract}

\end{frontmatter}

\section{Introduction}

Recently it was reported in \cite{kim2016robustness,panteley2017synchronization,jglee18automatica,lee2022design} that a network of heterogeneous agents exhibits an emergent behavior when the node dynamics are diffusively coupled with a high coupling gain. 
In particular, it turned out that the emergent behavior is governed by a so-called {\em blended dynamics}, which is simply the algebraic average of the node dynamics.
This finding yielded a methodology for designing heterogeneous agents that collectively perform a particular task or computation. 
That is, build a dynamic system first (as the blended dynamics) that performs the desired task, and then, assign to each node dynamics a piece of the vector field of the blended dynamics, so that the algebraic average of all the assigned vector fields becomes the blended dynamics.
This design method has been employed for, e.g., distributed optimization \cite{yun2019initialization}, distributed estimation of network size \cite{lee2018distributed}, distributed observer \cite{kim2019completely}, and so on.\\
However, all the aforementioned results use linear diffusive coupling for exchanging information among the agents, and the lower bound for the linear coupling gain depends on global information such as the number of participating agents, the graph structure, the vector fields of node dynamics, and so on.
Since this drawback restricts applicability of the method, the (node-wise) funnel coupling was studied in \cite{shim2015preliminary,lee2019synchronization}, which is a nonlinear coupling whose design does not depend on the global information.
Unfortunately, the introduction of the (node-wise) funnel coupling no longer yields the collective emergent behavior governed by the algebraic average of all the node dynamics (but by a nonlinear function of node dynamics given in an implicit form called the emergent dynamics; see \cite{lee2019synchronization} for more details).
This is a serious drawback, because no explicit expression for the emergent dynamics is available and hence the design of the dynamic system that governs the emergent behavior is no longer a simple task.\\
Inspired by the preliminary studies \cite{trenn2017edge,leeutility}, this paper presents an alternative to the (node-wise) funnel coupling in \cite{lee2019synchronization}, which is {\em edge-wise funnel coupling}. 
The advancement is that the funnel technique is directly applied to the output difference between two nodes connected by an edge.
In particular, we prove that edge-wise funnel coupling has the following benefits:
\begin{itemize}

\item The design of the edge-wise funnel coupling does not need global information. 

\item By designing the funnel, the convergence rate and the residual error of the output difference between two nodes can be arbitrarily controlled, which is an inherent advantage of funnel control (that was introduced in \cite{ilchmann2002tracking}; see \cite{BergLe18,BergIlch21} for recent advances).

\item The collective emergent behavior is governed by the blended dynamics (i.e., the simple algebraic average of the vector fields of the participating nodes when they have no internal dynamics; when they have internal dynamics, the blended dynamics take the form presented for the case of linear diffusive coupling in~\cite{jglee18automatica}).

\item During the operation of the multi-agent system, agents can leave and join the network without interrupting the operation of the system, which is called the `plug-and-play' property \cite{jglee18automatica}.
Edge-wise funnel coupling enables a simple handshaking procedure between two nodes of a newly created edge for the plug-and-play operation.

\end{itemize}
The paper is structured as follows.
Section 2 introduces a class of heterogeneous multi-agent systems considered in this paper, and presents the edge-wise funnel coupling.
Edge-wise funnel coupling employs the funnel function to every edge with the goal that the output difference of two nodes connected by the edge remains within the funnel.
This goal is called `funnel objective' in this paper, and is achieved in Section 2 under a symmetry assumption on the funnel functions and under an assumption that the blended dynamics has no finite escape time.
When the funnel objective is achieved and the funnel size shrinks to zero or to a small number as time goes by, the (asymptotic or approximate) output synchronization is achieved with a connected graph.
The question whether the funnel coupling remains bounded even when the funnel size shrinks to zero is also answered in Section 2.
Section 3 shows that, if the output synchronization is achieved and if the blended dynamics is stable in a certain sense, then an emergent behavior arises among the heterogeneous agents and the behavior is described by the solution to the blended dynamics.
A simulation result with the plug-and-play operation is illustrated in Section 4, and the conclusion follows in Section 5.\\
While the analysis for the funnel objective is motivated by \cite{lee2019synchronization}, in contrast to that work, our results take into account the internal dynamics of each node and are based on a new graph theoretical lemma, which quantifies the effect of consensus in each edge for arbitrarily given edge weights, and thus, is useful in the analysis of time-varying or state-dependent edge coupling gains like the edge-wise funnel coupling.
This lemma is found in the Appendix.
Finally, we remark that similar coupling laws have been presented in \cite{bechlioulis2014robust,bechlioulis2015robust,bechlioulis2016decentralized,verginis2019robust,macellari2016multi,mehdifar2020prescribed,stamouli2019robust}.
However, they either consider a leader-follower formulation~\cite{bechlioulis2014robust,bechlioulis2015robust,bechlioulis2016decentralized} (which corresponds to a tracking control problem), a specific graph structure, e.g., a tree graph~\cite{verginis2019robust} or infinitesimal rigidity~\cite{mehdifar2020prescribed} (which simplifies the analysis), or homogeneous agents~\cite{macellari2016multi,mehdifar2020prescribed} (which again simplifies the analysis). 
The problem of dynamic average consensus was solved in~\cite{stamouli2019robust} using a prescribed performance control (which shares some features of funnel control).
These works, however, do not consider the emergent behavior which is the focus of the current paper.

\section{Edge-wise funnel coupling law}\label{sec:pro}

In the present paper, we consider a heterogeneous multi-agent system given by 
\begin{align}\label{eq:eachdyn}
\begin{split}
\dot{\yy}_i(t) &= \FF_i(t,\yy_i(t), \zz_i(t)) + \boldsymbol{\Gamma}_i(t,\ww_i(t)) \cdot \uu_i(t),  \\
\dot{\zz}_i(t) &= \ZZ_i(t, \zz_i(t), \yy_i(t)), \\
\ww_i(t) &= \WW_i(\yy_i(t), \zz_i(t)), \,\quad\quad\quad\quad\quad\quad i \in \mathcal{N}.
\end{split}
\end{align}
Here, $\mathcal{N} := \{1, \dots, N\}$ is the set of agent indices, the number of agents is $N$, $\uu_i(t) \in \mathbb{R}^m$ is the coupling law to be designed, $\ww_i(t) \in \mathbb{R}^{m_i}$ is the introspective\footnote{This terminology was used in, e.g., \cite{grip2012output}, whose meaning is that the variable can be measured within the agent. 
We will use the value of $\ww_i$ when we compose the coupling law $\uu_i$.} output whose dimension $m_i$ may vary across the agents, $\yy_i(t) \in \mathbb{R}^m$ is the output with (agent-independent) dimension $m$ which is communicated with other agents and is to be synchronized approximately, and $\zz_i(t) \in \mathbb{R}^{n_i}$ is the internal state with (agent-dependent) dimension $n_i$.
The following two assumptions pose the required properties for $\FF_i$, $\ZZ_i$, $\WW_i$, and $\boldsymbol{\Gamma}_i$.

\begin{assum1}[Open loop dynamics]\label{assum:ind_vec_prop}
The functions $\FF_i: [t_0, \infty)\times \R^m \times \R^{n_i}\to\R^{m}$, $\ZZ_i:[t_0,\infty)\times\R^{n_i}\times\R^m\to\R^{n_i}$ and $\boldsymbol{\Gamma}_i:[t_0, \infty)\times \R^{m_i}\to\R^{m\times m}$ are measurable in~$t$, locally Lipschitz with respect to $(\yy_i, \zz_i)$ or $\ww_i$, resp., and bounded on each compact subset of $\mathbb{R}^{m+n_i}$ or $\R^{m_i}$, resp., uniformly in $t$. 
The function $\WW_i:\R^m\times\R^{n_i}\to\R^{m_i}$ is locally Lipschitz.
\end{assum1}

\begin{assum1}[Gain matrix]\label{assum:rel_deg}
The gain matrix $\boldsymbol{\Gamma}_i(t,\ww_i)$ is known and available for the design of the coupling law $\uu_i$, and is invertible for all $t$ and $\ww_i$.
Its inverse is uniformly bounded, i.e., there exists $M_{\boldsymbol{\Gamma}} > 0$ such that\footnote{The symbol $\|\cdot\|_\infty$ denotes the maximum norm for a vector and the induced maximum norm for a matrix.} $\|\boldsymbol{\Gamma}_i(t, \ww_i)^{-1}\|_\infty \le M_{\boldsymbol{\Gamma}}$ for all $t$ and $\ww_i$.
\end{assum1}

Note that Assumption~\ref{assum:rel_deg} justifies to say that system~\eqref{eq:eachdyn} has relative degree one, because for time-invariant systems the definition given in~\cite{ByrnIsid91a} is then satisfied.
Under the above assumptions, we propose for each $i \in \cN$ the \emph{edge-wise funnel coupling} law
\begin{align}\label{eq:funnel_coup}
\begin{split}
&\uu_i(t) \!=\! \uu_i\big(t,  \ww_i, \{\boldsymbol{\nu}_{ij}\}\big)\! = \! \boldsymbol{\Gamma}_i(t, \ww_i)^{-1}\!\!\sum_{j \in \mathcal{N}_i} \! \uu_{ij}(t, \boldsymbol{\nu}_{ij}), \\
&\uu_{ij}(t, \boldsymbol{\nu}_{ij}) = {\rm col}\left(\mu_{ij}^1\left(\tfrac{\nu_{ij}^1(t)}{\psi_{ij}^1(t)}\right), \dots, 
\mu_{ij}^m\left(\tfrac{\nu_{ij}^m(t)}{\psi_{ij}^m(t)}\right)\right), \\
&\boldsymbol{\nu}_{ij} := \yy_j - \yy_i = {\rm col}(\nu_{ij}^1, \dots, \nu_{ij}^m), 
\end{split}
\end{align}
where $\mathcal{N}_i \subset \mathcal{N}$ is the set of agents that send information to agent~$i$.
The communication graph and the functions $\psi_{ij}^p$ and $\mu_{ij}^p$ satisfy the following assumptions.

\begin{assum1}[Communication graph]\label{assum:graph}
The communication graph $\cG = (\cN, \cE)$ induced by the neighborhoods $\cN_i$ for $i\in\cN$ (i.e., $\cN$ is the set of nodes and $(j, i) \in \cE$ if, and only if, $j \in \cN_i$) is undirected and connected.\footnote{Different from the literature \cite{Dies17}, in the present paper edges $(j,i)\in\cE$ always have a direction (from node $j$ to node $i$), and a graph is undirected, if for any $(j,i)\in\cE$ we also have $(i,j)\in\cE$.}
\end{assum1}

For the basics of graph theory we refer to~\cite{Dies17}; some specific lemmas required for the proofs of the main results can also be found in Appendix~\ref{app:ess_graph_lem}.

\begin{assum1}[Design functions]\label{assum:functions}
For each edge $(i,j)\in\cE$ and $p\in\cM:=\{1,2,\ldots,m\}$ we have: 
\begin{itemize}

\item \emph{Performance functions} $\psi_{ij}^p: [t_0, \infty) \to \mathbb{R}_{> 0}$ are bounded and continuously differentiable with bounded derivatives.
Furthermore, they are symmetric in the sense that $\psi_{ij}^p(t) = \psi_{ji}^p(t)$, $\forall t \ge t_0$.

\item \emph{Coupling functions} $\mu_{ij}^p : (-1, 1) \to \mathbb{R}$ are continuous and satisfy $\lim_{s \to \pm 1} \mu_{ij}^p(s) = \pm\infty$. 
Furthermore, they satisfy the symmetry property that $\mu_{ij}^p(-s)=-\mu_{ji}^p(s)$, $\forall s\in(-1,1)$.
\end{itemize}
\end{assum1}

Under the simple coupling law $\uu_i$ as in \eqref{eq:funnel_coup} and the above assumptions we will prove that the following `funnel objective' is achieved: 
\begin{align}\label{eq:res_fun_edge}
\forall\, t \ge t_0\ \forall\, (j,i) \in \mathcal{E}\ \forall\, p\in \cM:\ \left|\nu_{ij}^p(t)\right| < \psi_{ij}^p(t)
\end{align}
(whose meaning is that the signal $\nu_{ij}^p$ remains inside the funnel characterized by $\psi_{ij}^p$).
While the choice of $\psi_{ij}^p$ is completely up to the designer, it is often chosen as a monotonically decreasing function (so that the funnel shrinks as time goes on).
By designing the function $\psi_{ij}^p$, one can control the upper bound of $|\nu_{ij}^p(t)|$ during the transient and the residual error $\limsup_{t \to \infty}|\nu_{ij}^p(t)|$.
Note that we allow for $\lim_{t\to\infty} \psi_{ij}^p(t) = 0$.\footnote{In this case, since $\lim_{t\to\infty} \nu_{ij}^p(t) \to 0$, the coupling law~\eqref{eq:funnel_coup} thus contains the quotient of two ``infinitesimally small'' terms. 
Therefore, the case of asymptotic synchronization seems to be of limited practical utility; similar to asymptotic tracking by funnel control, cf.~\cite[Rem.~1.7]{BergIlch21}.}

We also note that the design of $\uu_i$ in \eqref{eq:funnel_coup} does neither use the information of the vector fields $\FF_i$ and $\ZZ_i$ nor the state $\zz_i$.
When $\boldsymbol{\Gamma}_i(t, \ww_i)$ does not depend on $\ww_i$ (which is often the case, e.g., when $\boldsymbol{\Gamma}_i(t, \ww_i) = I$), the introspective output $\ww_i$ does not need to be measured.
Finally, we emphasize that the information of $\yy_j$ and $\yy_i$ themselves is not needed as long as the difference $\boldsymbol\nu_{ij}$ is available.
This is useful in some practical applications. 
For example, a self-driving car $i$ can easily measure the distance $y_j-y_i$ from the front car $j$, but the absolute positions $y_j$ and $y_i$ are hard to measure.

\begin{rem1}[Symmetry]
Note that the symmetry in the functions $\psi_{ij}^p$ and $\mu_{ij}^p$, stated in Assumption \ref{assum:functions}, are already assumed in the linear coupling law $\uu_i = \sum_{j \in \mathcal{N}_i} k \boldsymbol{\nu}_{ij}$, with a constant $k>0$, used in \cite{kim2016robustness,panteley2017synchronization,jglee18automatica,lee2022design}.
(Indeed, this is the case when $\mu_{ij}^p$ is the identity function and $\psi_{ij}^p \equiv 1$.)
Therefore, the edge-wise funnel coupling~\eqref{eq:funnel_coup} can be viewed as a generalization of these approaches in that, instead of the constant uniform gain~$k$, each edge has its individual nonlinear time-varying gain function.
\end{rem1}

\begin{rem1}[Normal form]
The proposed coupling law~\eqref{eq:funnel_coup} can be easily obtained even when the node dynamics is not in the normal form as in~\eqref{eq:eachdyn}.
For example, consider the node dynamics given by $\dot x_i = f_i(x_i) + g_i(x_i) u_i$ and $y_i = h_i(x_i)$.
If $L_{g_i}h_i(x_i) := h_i'(x_i) g_i(x_i)$ depends only on an introspective output $\ww_i$ and is nonsingular for all $\ww_i$ (i.e., the system has relative degree one~\cite{ByrnIsid91a}), then the coupling law~\eqref{eq:funnel_coup} can still be constructed with $\boldsymbol{\Gamma}_i(\ww_i) = L_{g_i}h_i(x_i)$.
\end{rem1}

Solutions to the differential equations of the closed-loop system~\eqref{eq:eachdyn} and~\eqref{eq:funnel_coup} are understood in the sense of Carath\'{e}odory, and their existence and uniqueness (local in time) follow from the assumptions.
Throughout the paper, when speaking of solutions, we will always mean the unique (maximal) Carath\'{e}odory solution.

Now, to guarantee \eqref{eq:res_fun_edge}, we need one more assumption: the blended dynamics (to be defined) has no finite escape time.
For this, define $\sss(t) := (1/N)\sum_{i=1}^N \yy_i(t)$. 
Then
\begin{align*}
\dot{\sss}(t) &= \frac{1}{N}\sum_{i=1}^N\left[ \FF_i(t, \yy_i(t), \zz_i(t)) + \boldsymbol{\Gamma}_i(t, \ww_i(t)) \cdot \uu_i(t) \right] \\
&= \frac{1}{N}\sum_{i=1}^N \FF_i(t, \yy_i(t), \zz_i(t))
\end{align*}
where the coupling terms cancel out because of the symmetry in Assumption \ref{assum:functions}.
Denoting the synchronization error by $\ee_i(t) := \yy_i(t) - \sss(t)$, we have
\begin{equation}\label{eq:blended_input}
\begin{aligned}
\dot{\sss}(t) &= \frac{1}{N}\sum_{i=1}^N \FF_i(t, \sss(t) + \ee_i(t), \zz_i(t)),\\
\dot{\zz}_i(t) &=  \ZZ_i(t, \zz_i(t), \sss(t) + \ee_i(t)), \quad\quad\quad\quad i \in \cN.
\end{aligned}
\end{equation}
When $\ee_i \equiv 0$ for all $i \in \mathcal{N}$, the system \eqref{eq:blended_input} is called the {\em blended dynamics} in~\cite{jglee18automatica}.
Here, we call~\eqref{eq:blended_input} the {\em perturbed blended dynamics} when we treat $\ee_i$, $i \in \cN$, as independent input signals to the blended dynamics.
In particular, we note that, if \eqref{eq:res_fun_edge} holds, then from Assumption~\ref{assum:graph}, we get
\vspace{-2mm}
\begin{align}\label{eq:sync_perf}
\forall\, t \ge t_0\ \forall\, i, j \in \mathcal{N}:\ \left\|\yy_i(t) - \yy_j(t)\right\|_\infty \le d_\cG \Psi(t),\vspace{-2mm}
\end{align}
where $d_\cG$ is the diameter\footnote{The diameter of a graph $\cG$ is the maximum length among the shortest paths between any two nodes.} of the communication graph $\cG = (\cN, \cE)$ and $\Psi(t) := \max_{p \in \cM}\max_{(j,i)\in \mathcal{E}} \psi_{ij}^p(t)$, and we find
\vspace{-2mm}
\begin{multline}\label{eq:key}
\hspace*{-3mm}\forall\, t \ge t_0\ \forall\, i\in\cN: \\
\hspace*{-2mm}\|\ee_i(t)\|_\infty = \left\| \tfrac{1}{N} \sum\nolimits_{j=1}^N (\yy_i(t)-\yy_j(t)) \right\|_\infty \le d_\cG \Psi(t).\vspace{-2mm}
\end{multline}

\begin{assum1}[No finite escape time]\label{assum:finite}
For any initial time $t_0$, the perturbed blended dynamics~\eqref{eq:blended_input} with any absolutely continuous inputs $\ee_i: [t_0, \infty) \to \mathbb{R}^m$, $i \in \mathcal{N}$, such that $\|\ee_i(t)\|_\infty \le d_\cG\Psi(t)$ for all $t \ge t_0$, has a global solution for any initial values $\sss(t_0)\in\R^m$, ${\zz}_i(t_0)\in\R^{n_i}$, $i \in \cN$.
\end{assum1}

We stress that if the functions $\FF_i$ and $\ZZ_i$ are globally Lipschitz in their arguments, then Assumption~\ref{assum:finite} holds.

\begin{lem1}\label{lem:finite}
Under Assumptions~\ref{assum:ind_vec_prop}--\ref{assum:finite}, assume that a solution of system \eqref{eq:eachdyn} with \eqref{eq:funnel_coup} exists on $[t_0, \omega)$ for some $\omega > t_0$ and satisfies $|\nu_{ij}^p(t)| < \psi_{ij}^p(t)$ for all $t \in [t_0, \omega)$, $(j,i) \in \mathcal{E}$, and $p \in \cM$.
Then $(\yy_i, \zz_i)$ is bounded on $[t_0, \omega)$ for all $i \in \cN$. 
\end{lem1}
The proof of Lemma~\ref{lem:finite} is a direct consequence of the representation~\eqref{eq:blended_input}, Assumption~\ref{assum:finite}, and~\eqref{eq:key}; the details are omitted.

\begin{thm}[Evolution inside funnel]\label{thm:funnel}
Consider the system~\eqref{eq:eachdyn} with the edge-wise funnel coupling law~\eqref{eq:funnel_coup}.
Under Assumptions \ref{assum:ind_vec_prop}--\ref{assum:finite}, if the initial values $\yy_i(t_0)$ of~\eqref{eq:eachdyn} and the performance functions $\psi_{ij}^p$ satisfy
\vspace{-2mm}
\begin{equation}\label{eq:lookatme}
\forall\, (j,i) \in \mathcal{E}\ \forall\, p \in \cM : \ |\nu_{ij}^p(t_0)| < \psi_{ij}^p(t_0),\vspace{-2mm}
\end{equation}
then the global solution $(\yy_1,\zz_1,\ldots,\yy_N,\zz_N):[t_0,\infty)\to\R^{Nm+n_1+\ldots+n_N}$ of~\eqref{eq:eachdyn} and~\eqref{eq:funnel_coup} exists, which satisfies the funnel objective~\eqref{eq:res_fun_edge}.
\end{thm}

The proof is relegated to Appendix~\ref{app:proof_diff}.
Note that, under the assumptions of Theorem~\ref{thm:funnel}, the inequality~\eqref{eq:sync_perf} holds, and thus, approximate (when $\limsup_{t \to \infty} \Psi(t) > 0$ is small) or asymptotic (when $\lim_{t \to \infty}\Psi(t) = 0$) output synchronization is achieved.

\begin{rem1}[Plug-and-play]\label{Rem:plug-and-play}
In virtue of Theorem~\ref{thm:funnel}, the multi-agent system~\eqref{eq:eachdyn} with the edge-wise funnel coupling~\eqref{eq:funnel_coup} is amenable to the plug-and-play operation; that is, agents can leave the network at any time (which, however, may decompose the network into several connected components), and agents can also join the network with no initialization of any agent in the network (see \cite{lee2022design} for more details).
In practice, both the required symmetry of $\psi_{ij}^p$ and $\mu_{ij}^p$ in Assumption \ref{assum:functions} and the condition \eqref{eq:lookatme} can be implemented when a new edge $(i,j)$ between agent $i$ and agent $j$ is created.
That is, the first communication between agent $i$ and agent $j$ is a handshake for these properties.
For example, under the premise that all the coupling functions are the same as $\mu_{ij}^p = s/(1-|s|)$ and the performance function has the form $\psi_{ij}^p(t) = \psi_{ji}^p(t) = (B^p - \eta) e^{-\lambda(t-t_k)} + \eta$ where $\eta$ and $\lambda$ are already determined, the undetermined $B^p$ and $t_k$ are negotiated by the handshake such that $t_k$ is set as the time of the handshake and $B^p$ is determined as $|y_i^p(t_k)-y_j^p(t_k)| < B^p$.
Then, all the conditions of Assumption \ref{assum:functions} hold and we have 
\vspace{-2mm}
\begin{equation}\label{eq:initialset}
\forall\, p \in \cM: \quad \left|\nu_{ij}^p(t_k)\right| < \psi_{ij}^p(t_k) = \psi_{ji}^p(t_k)\vspace{-2mm}
\end{equation}
so that the condition \eqref{eq:lookatme} holds at time $t_k$ and Theorem \ref{thm:funnel} applies afterward.
\end{rem1}

We note that, similar to~\cite[Rems.~1~\&~2]{lee2019synchronization}, the edge-wise funnel coupling law~\eqref{eq:funnel_coup} is also able to achieve finite-time synchronization.
Moreover, the coupling law \eqref{eq:funnel_coup} is guaranteed to remain bounded (even when the performance functions $\psi_{ij}^p$ converge to zero), under mild additional assumptions.

\begin{thm}[Boundedness of coupling law]\label{thm:bound_gain}
In addition to the assumptions of Theorem~\ref{thm:funnel}, assume that at least one of the following holds.
\begin{enumerate}[(a)]
\item $\FF_i(t, \yy, \zz) \equiv \widehat{\FF}(t, \yy) + \widetilde{\FF}_i(t, \yy, \zz)$, where $\widehat{\FF}(t, \yy)$ is globally Lipschitz with respect to $\yy$ uniformly in $t$ and there exists $M_{\FF}$ such that $\|\widetilde{\FF}_i(t, \yy, \zz)\|_\infty \le M_{\FF}$ for all $i \in \cN$, $t \ge t_0$, $\yy \in \mathbb{R}^m$, and $\zz \in \mathbb{R}^{n_i}$.
\item There exists $M_{\yy,\zz}$ such that $\|{\rm col}(\yy_i(t), \zz_i(t))\|_\infty \le {M}_{\yy, \zz}$ for all $i \in \cN$ and $t \ge t_0$.
\end{enumerate}
Then the input $\uu_i$ of \eqref{eq:funnel_coup} for \eqref{eq:eachdyn} is bounded on $[t_0, \infty)$, i.e., there exists ${M}_\uu > 0$ such that for all $t\ge t_0$ and $i \in \cN$, we have $\|\uu_i(t)\|_\infty \le {M}_\uu$. 
\end{thm}
The proof is similar to that of~\cite[Thm.~3]{lee2019synchronization}, when we additionally invoke the boundedness of $\boldsymbol{\Gamma}_i^{-1}$ from Assumption~\ref{assum:rel_deg}; hence it is omitted. 

\section{Blended dynamics as emergent behavior}\label{sec:spe}

Theorem \ref{thm:funnel} in the previous section provides sufficient conditions for the funnel objective~\eqref{eq:res_fun_edge} to be achieved.
In this section, we show that, if the funnel shrinks (i.e., $\Psi(t)$ gets small) as time goes by, then an emergent behavior arises which is described by the solution to the blended dynamics.
In fact, this emergence is based on a certain stability of the blended dynamics, and, in this section, we utilize ISS (input-to-state stability) and $\delta$-ISS (incremental ISS) of the perturbed blended dynamics~\eqref{eq:blended_input}, which are briefly reviewed in the following subsection.

\subsection{ISS and $\delta$-ISS}

Consider a system $\dot{\xx}(t) = \FF(t, \xx(t), \uu(t))$ whose solution exists globally in time for any initial condition $\xx(t_0)$ and for any locally essentially bounded measurable input $\uu$.
The system is {\em ISS with $(\beta,\hat\gamma)$ for a closed set $A$}, if there exist $\beta \in \KL$ and $\hat\gamma \in \Kinf$, such that, for all $t \ge t_0$,\footnote{For a set $\Xi \subseteq \R^n$, $\|\xx\|_{\Xi}$ denotes the distance between the point $\xx \in \R^n$ and $\Xi$, i.e., $\| \xx\|_{\Xi} := \inf_{\yy \in \Xi} \| \xx -\yy\|_\infty$.}
\begin{equation}\label{eq:siss}
\left\|\xx(t)\right\|_A \le \beta(\| \xx(t_0) \|_A, t- t_0)
+ \hat{\gamma}({\textstyle\sup}_{s \in [t_0, t)} \|\uu(s)\|_\infty).
\end{equation}
On the other hand, the system is {\em $\delta$-ISS with $(\beta,\hat\gamma)$} if there exist $\beta \in \KL$ and $\hat\gamma \in \Kinf$ such that, for any $\hat{\xx}(t_0)$ and $\xx(t_0)$ and for any locally essentially bounded measurable inputs $\hat{\uu}$ and $\uu$, the corresponding solutions $\hat{\xx}$ and $\xx$, respectively, satisfy, for all $t \ge t_0$,
\vspace{-2mm}
\begin{multline}\label{eq:sdeltaiss}
\|\hat{\xx}(t) - \xx(t)\|_\infty \le \beta(\|\hat{\xx}(t_0) - \xx(t_0)\|_\infty, t-t_0) \\
+ \hat{\gamma}\left(\begin{matrix}\sup_{s \in [t_0, t)}\end{matrix}\|\hat{\uu}(s) - \uu(s)\|_\infty\right).\vspace{-2mm}
\end{multline}

From \eqref{eq:siss} and \eqref{eq:sdeltaiss}, it follows from causality that, if the input $\uu$ (or, the input difference $\hat \uu-\uu$, resp.) converges to zero, then the state $\xx$ (or, the state difference $\hat \xx - \xx$, resp.) tends to zero.
However, to quantify an explicit rate of convergence for decaying inputs, we present a lemma whose proof is in Appendix~\ref{app:SecSpe}.

\begin{lem1}\label{lem:2}
If the system is ISS with $(\beta,\hat\gamma)$ for a closed set~$A$, then, for any $M_{\xx_0}>0$, $M_{\uu}>0$ and for any decreasing function $w:\R_{\ge 0}\to (0,1]$ such that $\lim_{t\to\infty}w(t)=0$, there exists $\gamma \in \Kinf$\footnote{An example of $\gamma$ in the explicit form is given in the proof of Lemma~\ref{lem:2}.} such that
\vspace{-2mm}
\begin{multline}\label{eq:sclaim}
\|\xx(t)\|_A \le \beta(\|\xx(t_0)\|_A,t-t_0) + \gamma(\delta_t), \quad t > t_0\vspace{-2mm}
\end{multline}
where
$$\delta_t := \sup_{s \in [t_0, t)} \|\uu(s)\|_\infty w(t-s),$$
for any solution $\xx$ with an initial condition $\xx(t_0)$ and a locally essentially bounded measurable input $\uu$ such that
\vspace{-2mm}
\begin{equation}\label{eq:scondi}
\|\xx(t_0)\|_A \le M_{\xx_0} \quad \text{and} \quad
\sup_{s \in [t_0,t)}\|\uu(s)\|_\infty \le M_{\uu}.\vspace{-2mm}
\end{equation}
\end{lem1}

We note that, for the system considered in this subsection, $\delta$-ISS with $(\beta, \hat{\gamma})$ of the system is equivalent to ISS with $(\beta, \hat{\gamma})$ for the closed set $A = \mathbb{R}^n \times \{0\}$ of the extended dynamics
\vspace{-2mm}
\begin{align*}
\dot{\hat{\xx}}(t) &= \FF(t, \hat{\xx}(t), \hat{\uu}(t)), \\
\dot{\ee}_\xx(t) &= \FF(t, \ee_\xx(t) + \hat{\xx}(t), \ee_\uu(t) + \hat{\uu}(t)) - \FF(t, \hat{\xx}(t), \hat{\uu}(t))\vspace{-2mm}
\end{align*}
with $\ee_\xx := \xx - \hat{\xx}$, input $\ee_\uu:=\uu - \hat{\uu}$ and for any signal~$\hat \uu$.
Therefore, Lemma \ref{lem:2} also applies to the $\delta$-ISS case.

\subsection{Emergent Behavior}

\begin{thm}[Emergent behavior]\label{thm:compact}
Let the assumptions of Theorem~\ref{thm:funnel} hold and assume that the perturbed blended dynamics~\eqref{eq:blended_input} is ISS with $(\beta, \hat{\gamma})$ for a closed set~$A$.
Then, for any $M_{\xx_0} > 0$ and any decreasing function $w:[0,\infty) \to (0,1]$ such that $\lim_{t\to\infty}w(t)=0$, there exists $\gamma \in \Kinf$ such that
\begin{align*}
&\max_{i \in \cN} \left\|\begin{bmatrix} \yy_i(t) \\ \zz_1(t) \\ \vdots \\ \zz_N(t) \end{bmatrix} \right\|_A \le \beta\left(\left\|\begin{bmatrix} \frac1N \sum_{i=1}^N \yy_i(t_0) \\ \zz_1(t_0) \\ \vdots \\ \zz_N(t_0) \end{bmatrix} \right\|_A, t - t_0\right) \\
&\qquad + \gamma(\begin{matrix}\sup_{s \in [t_0, t)}\end{matrix}d_\mathcal{G}\Psi(s)w(t-s)) + d_\mathcal{G}\Psi(t), \;\;\; \forall t \ge t_0
\end{align*}
for any solution ${\rm col}(\yy_1, \zz_1, \dots, \yy_N, \zz_N)$ of \eqref{eq:eachdyn} and \eqref{eq:funnel_coup} such that $\|{\rm col}((1/N) \sum_{i=1}^N \yy_i(t_0), \zz_1(t_0), \dots, \zz_N(t_0) )\|_A \le M_{\xx_0}$.
\end{thm}

\begin{pf}
Define $\yy_i^\zz := {\rm col}(\yy_i, \zz_1, \dots, \zz_N)$, for convenience, where $\yy_i$ and $\zz_i$, $i \in \cN$, are the solutions to~\eqref{eq:eachdyn} with~\eqref{eq:funnel_coup} such that $\|(1/N) \sum_{i=1}^N \yy_i^{\zz}(t_0)\|_A \le M_{\xx_0}$.
Now let $\sss^\zz := {\rm col}(\sss, \bar\zz_1, \dots, \bar\zz_N)$ be the solution to the perturbed blended dynamics \eqref{eq:blended_input} with the particular input $\ee_i = \yy_i - \sss$, $i \in \cN$ and the initial condition $\sss(t_0) = (1/N)\sum_{i=1}^N \yy_i(t_0)$ and $\bar{\zz}_i(t_0) = \zz_i(t_0)$, $i \in \mathcal{N}$.
It is seen from \eqref{eq:blended_input} that $\bar\zz_i(t) = \zz_i(t)$ for all $t \ge t_0$ with this particular input, and \eqref{eq:key} holds from Theorem~\ref{thm:funnel}.
Then, for each $i \in \cN$,
\begin{align*}
&\|\yy_i^\zz(t)\|_A \le \|\sss^\zz(t)\|_A + \| \yy_i(t) - \sss(t) \|_\infty \\
&\le \beta(\|\sss^\zz(t_0)\|_A, t - t_0) + \hat{\gamma}({\textstyle\sup}_{s \in [t_0, t)} d_\mathcal{G}\Psi(s)) + d_\mathcal{G}\Psi(t).
\end{align*}
Applying Lemma \ref{lem:2} with $M_\uu = d_\mathcal{G} \sup_{t \in [t_0,\infty)}\Psi(t)$, the proof is complete.
\QEDB
\end{pf}

When the perturbed blended dynamics \eqref{eq:blended_input} is ISS for a closed set $A$, the behavior of the {\em blended dynamics} (which is \eqref{eq:blended_input} with $\ee_i(t) \equiv 0$, $i \in \cN$):
\begin{align}\label{eq:blended}
\begin{split}
\dot{\hat{\sss}}(t) &= \frac{1}{N}\sum_{i=1}^N \FF_i(t, \hat{\sss}(t), \hat{\zz}_i(t)),\\
\dot{\hat{\zz}}_i(t) &=  \ZZ_i(t, \hat{\zz}_i(t), \hat{\sss}(t)), \quad\quad\quad\quad i \in \cN, 
\end{split}
\end{align}
is that all the solutions converge to the set $A$.
This is an emergent behavior because individual node dynamics do not necessarily have such a property.
With the edge-wise funnel coupling whose funnel shrinks such that $\limsup_{t\to\infty}\Psi(t)$ is sufficiently small or even zero, it is seen from Theorem \ref{thm:compact} that the behavior of \eqref{eq:eachdyn} with \eqref{eq:funnel_coup} mimics that of the blended dynamics.

On the other hand, even when the perturbed blended dynamics \eqref{eq:blended_input} do not have such an attractive set $A$, a similar phenomenon is observed if \eqref{eq:blended_input} is $\delta$-ISS.

\begin{cor1}[Emergent behavior]\label{cor:asymp}
Let the assumptions of Theorem~\ref{thm:funnel} hold and assume that \eqref{eq:blended_input} is $\delta$-ISS with $(\beta, \hat{\gamma})$.
Then, for any decreasing function $w:\R_{\ge 0}\to (0,1]$ such that $\lim_{t\to\infty}w(t)=0$, there exists $\gamma \in \Kinf$ such that, for each $i \in \cN$,
\begin{multline}\label{eq:error_blend}
\left\|\begin{bmatrix} \yy_i(t) \\ \zz_i(t) \end{bmatrix} - \begin{bmatrix} \hat{\sss}(t) \\ \hat{\zz}_i(t) \end{bmatrix} \right\|_\infty \\
\le \gamma({\textstyle\sup}_{s \in [t_0, t)} d_\mathcal{G} \Psi(s) w(t-s)) + d_\mathcal{G} \Psi(t),
\end{multline}
where $\yy_i$ and $\zz_i$, $i \in \cN$, are a solution to \eqref{eq:eachdyn} with \eqref{eq:funnel_coup}, and $\hat\sss$ and $\hat\zz_i$ are a solution to the blended dynamics \eqref{eq:blended} with $\hat\sss(t_0) = (1/N)\sum_{i=1}^N \yy_i(t_0)$ and $\hat\zz_i(t_0)=\zz_i(t_0)$, $i \in \cN$.
\end{cor1}

\begin{pf}
As in the proof of Theorem \ref{thm:compact}, let $\sss^\zz := {\rm col}(\sss, \bar\zz_1, \dots, \bar\zz_N)$ be the solution to the perturbed blended dynamics \eqref{eq:blended_input} with the particular input $\ee_i = \yy_i - \sss$, $i \in \cN$ and the initial condition $\sss(t_0) = (1/N)\sum_{i=1}^N \yy_i(t_0)$ and $\bar{\zz}_i(t_0) = \zz_i(t_0)$, $i \in \mathcal{N}$, so that $\bar\zz_i(t) = \zz_i(t)$ for all $t \ge t_0$, and \eqref{eq:key} holds.
Then, for each $i \in \cN$,
\vspace{-2mm}
\begin{align*}
&\left\|\begin{bmatrix} \yy_i(t) \\ \zz_i(t) \end{bmatrix} - \begin{bmatrix} \hat{\sss}(t) \\ \hat{\zz}_i(t) \end{bmatrix} \right\|_\infty 
\le \left\|\begin{bmatrix} \yy_i(t) \\ \zz_i(t) \end{bmatrix} - \begin{bmatrix} \sss(t) \\ \bar{\zz}_i(t) \end{bmatrix} \right\|_\infty \\
&+ \left\|\begin{bmatrix} \sss(t) \\ \bar\zz_i(t) \end{bmatrix} - \begin{bmatrix} \hat{\sss}(t) \\ \hat{\zz}_i(t) \end{bmatrix} \right\|_\infty 
\le d_\mathcal{G} \Psi(t) + \hat\gamma\left(\sup_{s \in [t_0, t)} d_\mathcal{G} \Psi(s)\right)\vspace{-2mm}
\end{align*}
which follows from $\delta$-ISS of the perturbed blended dynamics, and the fact that $\sss(t_0)=\hat\sss(t_0)$ and $\bar\zz_i(t_0)=\hat\zz_i(t_0)$, $i\in\cN$.
Finally, applying Lemma \ref{lem:2} with any $M_{\xx_0}>0$ and $M_\uu = d_\mathcal{G} \sup_{t \in [t_0,\infty)}\Psi(t)$, the proof is complete.
\QEDB
\end{pf}

Again, inequality \eqref{eq:error_blend} shows how the individual solution $\yy_i$ and $\zz_i$ is approximated by the solution to the blended dynamics \eqref{eq:blended} when the funnel shrinks as time goes by.
In particular, if asymptotic output consensus is achieved with $\lim_{t\to \infty}\Psi(t)=0$, then~\eqref{eq:error_blend} implies that the behavior of the network~\eqref{eq:eachdyn} with~\eqref{eq:funnel_coup} asymptotically tends to the behavior of the blended dynamics~\eqref{eq:blended}.

\begin{rem1}\label{rem:emerg}
Let the assumptions of Corollary~\ref{cor:asymp} hold.
Now, assume that there exists at least one bounded solution ${\rm col}(\sss, \zz_1, \dots, \zz_N)$ of the perturbed blended dynamics~\eqref{eq:blended_input} with some bounded input $\ee_i$, $i \in \mathcal{N}$.
Then, $\delta$-ISS of~\eqref{eq:blended_input} implies that any solution of~\eqref{eq:blended_input} with bounded input is bounded.
In particular, any solution ${\rm col}(\hat{\sss}, \hat{\zz}_1, \dots, \hat{\zz}_N)$ of the blended dynamics~\eqref{eq:blended} is bounded (since $\ee_i \equiv 0$, $i \in \mathcal{N}$).
This further implies by~\eqref{eq:error_blend} that any solution ${\rm col}(\yy_1, \zz_1, \dots, \yy_N, \zz_N)$ of~\eqref{eq:eachdyn} with~\eqref{eq:funnel_coup} is bounded, which gives that the corresponding inputs $\uu_i$, $i\in\cN$, are bounded by Theorem~\ref{thm:bound_gain}.
\end{rem1}

\subsection{Identical internal dynamics}

In the remainder of this section we consider the special case when all the internal dynamics (the differential equation for $\zz_i$) share the same vector field, i.e., $\ZZ_i = \ZZ$ for all $i\in\cN$, but not necessarily the same initial condition.
In this case, it may be convenient to consider a {\em reduced order blended dynamics}:
\vspace{-2mm}
\begin{subequations}\label{eq:red_blend_full}
\begin{align}
\dot{\tilde{\sss}}(t) &= \frac{1}{N}\sum_{i=1}^N \FF_i(t, \tilde{\sss}(t), \tilde{\zz}(t)), \label{eq:red_blend} \\
\dot{\tilde{\zz}}(t) &= \ZZ(t, \tilde{\zz}(t), \tilde{\sss}(t)).\vspace{-2mm} \label{eq:nom_z_dyn}
\end{align}
\end{subequations}
This is motivated by the observation that, under the assumption that $\ZZ_i = \ZZ$, $i \in \mathcal{N}$, if the perturbed blended dynamics~\eqref{eq:blended_input} is $\delta$-ISS with $(\beta,\hat\gamma)$, then the set 
\begin{align*}\label{eq:sync_man}
\mathcal{S}^\zz := \setdef{{\rm col}(\hat{\sss}, \hat{\zz}_1, \dots, \hat{\zz}_N)}{ \forall\, i, j \in \mathcal{N}:\ \hat{\zz}_i = \hat{\zz}_j}
\end{align*}
is globally asymptotically stable for \eqref{eq:blended} (i.e., \eqref{eq:blended_input} with zero inputs) hence $\mathcal{S}^\zz$ is an invariant set of \eqref{eq:blended}.
Indeed, let ${\rm col}(\hat\sss, \hat\zz_1,\ldots,\hat\zz_N)$ be a solution of~\eqref{eq:blended}, then it also solves~\eqref{eq:blended_input} with $\ee_i \equiv 0$, $i\in\cN$.
Now, let $(\tilde{\sss}, \tilde{\zz})$ be any solution of~\eqref{eq:red_blend_full} such that $\tilde{\sss}(t_0) = \hat\sss(t_0)$, then ${\rm col}(\tilde{\sss}, \tilde{\zz},\ldots,\tilde{\zz})$ also solves~\eqref{eq:blended_input} with $\ee_i \equiv 0$, $i\in\cN$.
Therefore, for all $i \in \cN$ and for all $t \ge t_0$,
\begin{equation}\label{eq:dISStiS}
\left\|\begin{bmatrix} \hat \sss(t) \\ \hat \zz_i(t) \end{bmatrix} - \begin{bmatrix} \tilde\sss(t) \\ \tilde\zz(t) \end{bmatrix} \right\|_\infty \le \beta(\max_{j} \|\hat{\zz}_j(t_0) - \tilde{\zz}(t_0)\|_\infty, t - t_0).
\end{equation}

\begin{cor1}[Emergent behavior]\label{cor:red1}
Let the assumptions of Theorem~\ref{thm:funnel} hold and assume that \eqref{eq:blended_input} is $\delta$-ISS with $(\beta, \hat{\gamma})$.
Then, for any decreasing function $w:\R_{\ge 0}\to (0,1]$ such that $\lim_{t\to\infty}w(t)=0$, there exists $\gamma \in \Kinf$ such that, for each $i \in \cN$,
\vspace{-2mm}
\begin{align*}
&\left\|\begin{bmatrix} \yy_i(t) \\ \zz_i(t)\end{bmatrix} - \begin{bmatrix} \tilde{\sss}(t) \\ \tilde{\zz}(t) \end{bmatrix}\right\|_\infty \le \beta(\max_j \|\zz_j(t_0) - \tilde \zz(t_0)\|_\infty, t - t_0) \\
&\qquad\qquad + \gamma(\begin{matrix}\sup_{s \in [t_0, t)}\end{matrix}d_\mathcal{G}\Psi(s)w(t-s)) + d_\mathcal{G}\Psi(t)\vspace{-2mm}
\end{align*}
where $\yy_i$ and $\zz_i$, $i \in \cN$, are a solution to \eqref{eq:eachdyn} with \eqref{eq:funnel_coup}, and $\tilde\sss$ and $\tilde\zz$ are a solution to the reduced order blended dynamics \eqref{eq:red_blend_full} with $\tilde\sss(t_0) = (1/N)\sum_{i=1}^N \yy_i(t_0)$.
\end{cor1}

\begin{pf}
The claim follows from
\begin{align*}
&\left\|\begin{bmatrix} \yy_i(t) \\ \zz_i(t) \end{bmatrix} - \begin{bmatrix} \tilde{\sss}(t) \\ \tilde{\zz}(t) \end{bmatrix} \right\|_\infty 
\le \\
&\qquad \left\|\begin{bmatrix} \yy_i(t) \\ \zz_i(t) \end{bmatrix} - \begin{bmatrix} \hat{\sss}(t) \\ \hat{\zz}_i(t) \end{bmatrix} \right\|_\infty 
+ \left\|\begin{bmatrix} \hat \sss(t) \\ \hat\zz_i(t) \end{bmatrix} - \begin{bmatrix} \tilde{\sss}(t) \\ \tilde{\zz}(t) \end{bmatrix} \right\|_\infty 
\end{align*}
combined with \eqref{eq:error_blend} and \eqref{eq:dISStiS} and with $\hat\sss(t_0) = \tilde\sss(t_0)$ and $\hat\zz_i(t_0)=\zz_i(t_0)$, $i \in \cN$.
\QEDB
\end{pf}

While the solutions $\yy_i$ and $\zz_i$ are compared to the solution to the blended dynamics \eqref{eq:blended} in Corollary \ref{cor:asymp}, they are compared to the solution to the reduced order blended dynamics \eqref{eq:red_blend_full} in Corollary \ref{cor:red1}, which is simpler to compute.
However, the drawback is the addition of the transient term $\beta$, which is caused by the difference in the initial values $\zz_i(t_0)$ and $\zz_j(t_0)$, $i,j \in \mathcal{N}$.

Now, to best utilize the benefit of low dimensionality of the reduced order blended dynamics~\eqref{eq:red_blend_full}, it is desired that the stability condition is imposed on the reduced order blended dynamics \eqref{eq:red_blend_full} rather than \eqref{eq:blended_input}.
For this, let us introduce the {\em perturbed reduced order blended dynamics} with input ${\rm col}(\ee_0, \ee_1, \dots, \ee_N, \dd_1, \dots, \dd_N)$:
\vspace{-2mm}
\begin{align}\label{eq:red_blen_inp}
\begin{split}
\dot{\bar\sss}(t) &= \frac{1}{N}\sum_{i=1}^N \FF_i(t, \bar\sss(t) + \ee_i(t), \bar\zz(t) + \dd_i(t)), \\
\dot{\bar\zz}(t) &= \ZZ(t, \bar\zz(t), \bar\sss(t) + \ee_0(t)). \vspace{-2mm}
\end{split}
\end{align}
And, instead of $\delta$-ISS of \eqref{eq:blended_input}, we will assume $\delta$-ISS of \eqref{eq:red_blen_inp} with the additional assumption that the internal dynamics~\eqref{eq:nom_z_dyn} is also $\delta$-ISS with $(\tilde\beta,\bar\gamma)$ when $\tilde\sss$ is viewed as an input.
Indeed, in this case, the solutions $\zz_i$ and $\yy_i$, $i \in \cN$, of \eqref{eq:eachdyn} with \eqref{eq:funnel_coup} have the property that $\|\zz_j(t)-\zz_i(t)\|_\infty \le \tilde\beta(\|\zz_j(t_0)-\zz_i(t_0)\|_\infty,t-t_0)+\bar\gamma(\sup_{s\in[t_0,t)} \|\yy_j(s)-\yy_i(s)\|_\infty)$ for all $j,i \in \cN$.
If the funnel objective is achieved, then by \eqref{eq:sync_perf} and Lemma~\ref{lem:2}, for any $M_{\xx_0} > 0$ and any decreasing function $\tilde{w} : [0, \infty) \to (0, 1]$ such that $\lim_{t \to \infty} \tilde{w}(t) = 0$, there is $\tilde\gamma \in \Kinf$ such that, for any $j,i \in \mathcal{N}$,
\vspace{-2mm}
\begin{align*}
&\|\zz_j(t) - \zz_i(t)\|_\infty \le \max_{j,i} \tilde{\beta}(\|\zz_j(t_0) - \zz_i(t_0)\|_\infty, t  - t_0) \\
&\qquad\qquad\quad + \tilde{\gamma}(\begin{matrix} \sup_{s \in [t_0, t)}\end{matrix} d_\mathcal{G} \Psi(s)\tilde{w}(t-s)) =: D(t),\vspace{-2mm}
\end{align*}
for the solutions such that $\max_{j,i} \|\zz_j(t_0) - \zz_i(t_0)\|_\infty \le M_{\xx_0}$.
It is clear that $D(t)$ has similar properties as $\Psi(t)$; i.e., $\limsup_{t\to\infty}D(t)$ is small or zero if $\limsup_{t\to\infty}\Psi(t)$ is small or zero, resp.

\begin{cor1}[Emergent behavior]\label{thm:red}
Let the assumptions of Theorem~\ref{thm:funnel} hold and assume that the internal dynamics~\eqref{eq:nom_z_dyn} with input~$\tilde\sss$ is $\delta$-ISS.
If the perturbed reduced order blended dynamics~\eqref{eq:red_blen_inp} is $\delta$-ISS with $(\beta, \hat{\gamma})$, then for any decreasing function $w : [0, \infty) \to (0, 1]$ such that $\lim_{t \to \infty} w(t) = 0$, there exists $\gamma \in \Kinf$ such that, for each $i \in \mathcal{N}$,
\vspace{-2mm}
\begin{align*}
\begin{split}
&\left\|\begin{bmatrix} \yy_i(t) \\ \zz_i(t)\end{bmatrix} - \begin{bmatrix} \tilde{\sss}(t)\\ \tilde{\zz}(t)\end{bmatrix}\right\|_\infty \le \beta(\|\zz_i(t_0) - \tilde\zz(t_0)\|_\infty, t-t_0) \\
&\; + \gamma({\textstyle\sup}_{s \in [t_0, t)} \max\{\d_\mathcal{G}\Psi(s), D(s)\} w(t-s)) + d_\cG\Psi(t)\vspace{-2mm}
\end{split}
\end{align*}
where $\zz_i$ and $\yy_i$ are solutions to \eqref{eq:eachdyn} with \eqref{eq:funnel_coup} such that $\max_{j,i} \|\zz_j(t_0) - \zz_i(t_0)\|_\infty \le M_{\xx_0}$, and ${\rm col}(\tilde{\sss}, \tilde{\zz})$ is the solution of the reduced order blended dynamics~\eqref{eq:red_blend_full} with initial condition $\tilde{\sss}(t_0) = (1/N)\sum_{i=1}^N \yy_i(t_0)$ and $\tilde{\zz}(t_0) = (1/N)\sum_{i=1}^N \zz_i(t_0)$.
\end{cor1}

\begin{pf}
Pick any $i\in\cN$.
Then, any solution ${\rm col}(\yy_1, \zz_1, \dots, \yy_N, \zz_N)$ of~\eqref{eq:eachdyn} and~\eqref{eq:funnel_coup} satisfies
\vspace{-2mm}
\begin{align}\label{eq:szi}
\begin{split}
\dot{\sss}(t) &= \frac{1}{N}\sum_{j=1}^N \FF_j\big(t, \sss(t) + (\yy_j(t)-\sss(t)), \zz_i(t) \\
&\,\,\quad\quad\quad\quad\quad\quad\quad\quad\quad\quad\quad + (\zz_j(t)-\zz_i(t))\big) \\
\dot{\zz}_i(t) &= \ZZ(t, \zz_i(t), \sss(t) + (\yy_i(t)-\sss(t))) \vspace{-2mm}
\end{split}
\end{align}
where $\sss = (1/N)\sum_{j=1}^N \yy_j$.
Comparing \eqref{eq:szi} and \eqref{eq:red_blend_full} under $\delta$-ISS of \eqref{eq:red_blen_inp} with $\ee_0 = \yy_i - \sss$, $\ee_j = \yy_j - \sss$, and $\dd_j = \zz_j - \zz_i$,
\vspace{-2mm}
\begin{align*}
&\left\|\begin{bmatrix} \yy_i(t) \\ \zz_i(t)\end{bmatrix} - \begin{bmatrix} \tilde{\sss}(t)\\ \tilde{\zz}(t)\end{bmatrix}\right\|_\infty
\le \left\|\ee_i(t)\right\|_\infty + \left\|\begin{bmatrix} \sss(t) \\ \zz_i(t)\end{bmatrix} - \begin{bmatrix} \tilde{\sss}(t)\\ \tilde{\zz}(t)\end{bmatrix}\right\|_\infty \\
&\le \left\|\ee_i(t)\right\|_\infty + \beta(\| \zz_i(t_0) - \tilde{\zz}(t_0)\|_\infty, t-t_0) \\
&\quad + \hat\gamma\left(\sup_{s \in [t_0, t)}\left\|\begin{bmatrix}\|\ee_0(s)\|_\infty \\ \max_{j} \|\ee_j(s)\|_\infty \\ \max_{j} \|\dd_j(s)\|_\infty\end{bmatrix}\right\|_\infty\right).\vspace{-2mm}
\end{align*}
The proof concludes by applying Lemma \ref{lem:2} and by \eqref{eq:key}.
\QEDB
\end{pf}

\begin{rem1}
If \eqref{eq:red_blen_inp} is $\delta$-ISS with $(\beta,\hat\gamma)$ and the function $s - \hat\gamma(2s)$ is of class $\Kinf$, then it can be shown that \eqref{eq:blended_input} is $\delta$-ISS (whose proof is in the extended arXiv version of this paper).
Then, Corollary \ref{cor:red1} can be employed instead of Corollary \ref{thm:red}.
\end{rem1}

\section{An example}\label{sec:sim}

\begin{figure}[hbt]
\begin{center}
\includegraphics[width=0.8\columnwidth]{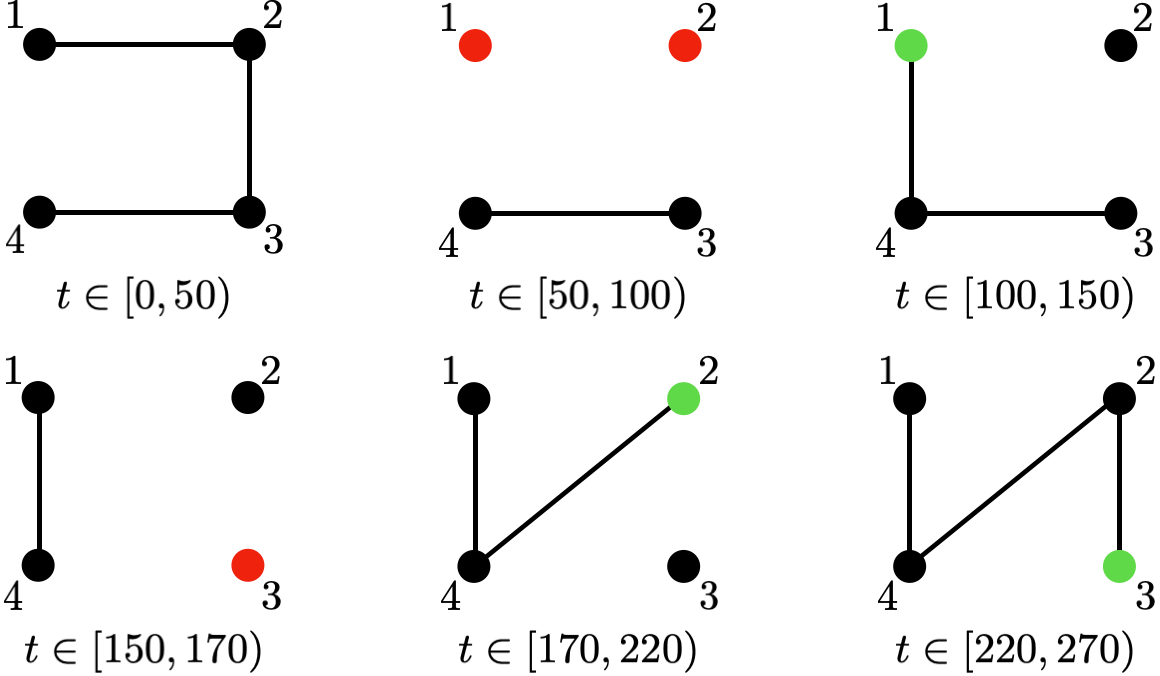}
\caption{Switched graph $\cG(t)$ and time intervals used for simulation.
Red dots are agents leaving the network. Green dots are agents newly joined.}
\label{fig:graph}
\end{center}
\end{figure}

In this section, we illustrate that different emergent behaviors can arise when some heterogeneous agents join or leave the network (plug-and-play operation).
Consider a system of four heterogeneous agents:
\vspace{-2mm}
\begin{align*}
\dot{y}_i(t) &= F_i(t, y_i(t), \zz_i(t)) + 100\, u_i(t), \\
\dot{\zz}_i(t) &= \ZZ(t, \zz_i(t), y_i(t)), \qquad\qquad\qquad i\!\in\!\cN \!=\! \{1,\ldots,4\},\vspace{-2mm}
\end{align*}
where $\zz_i(t) := {\rm col}(z_{i,1}(t), z_{i,2}(t), z_{i,3}(t))$, $\ZZ(t, \zz_i, y_i) := {\rm col}(Z_1(t, z_{i,1},y_i), Z_2(t, z_{i,2}, y_i), Z_3(t, z_{i,3}, y_i))$, and \vspace{-2mm}
\begin{align*}
Z_1(t, z, y) &:= -100z + 100y,\\
Z_2(t, z, y) &:= \begin{cases} -z + 0.4(y + 0.5), &\mbox{ if } y + 0.5 < 0, \\ -z + 7(y + 0.5), &\mbox{ if } y + 0.5 \ge 0, \end{cases} \\
Z_3(t, z, y) &:= \frac{1}{20}\begin{cases} -z, &\mbox{ if } y + 1 < 0, \\ -z + 50(y+1), &\mbox{ if } y+1 \ge 0.\end{cases}\vspace{-2mm}
\end{align*}
The heterogeneous vector fields $F_i$ are given by
\vspace{-2mm}
\begin{align*}
F_1(t, y_1, \zz_1) &:= -\tfrac{100}{3}y_1^3 + 400 z_{1,1} + 1100, \\
F_2(t, y_2, \zz_2) &:= -\tfrac{100}{3}y_2^3 - 1600 z_{2,2} - \tfrac{5500}{3},\\
F_3(t, y_3, \zz_3) &:= -\tfrac{100}{3}y_3^3 \!+\! 1600 z_{3,2} \!-\! (20z_{3,2} \!-\! 22)^2 \!+\! \tfrac{1100}{3}, \\
F_4(t, y_4, \zz_4) &:= -\tfrac{100}{3}y_4^3 - 400 z_{4,3} + \tfrac{5500}{3}.\vspace{-2mm}
\end{align*}
Each agent represents a neuromorphic circuit with one positive/negative feedback inspired by~\cite{ribar2019neuromodulation}.
Since $\Gamma_i \equiv 100$ for all agents, they satisfy Assumptions~\ref{assum:ind_vec_prop} and~\ref{assum:rel_deg}.
Fig.~\ref{fig:graph} illustrates the switched graph $\cG(t)$ with the corresponding time intervals for the demonstration.

\begin{figure}[hbt]
\begin{center}
\includegraphics[width=\columnwidth]{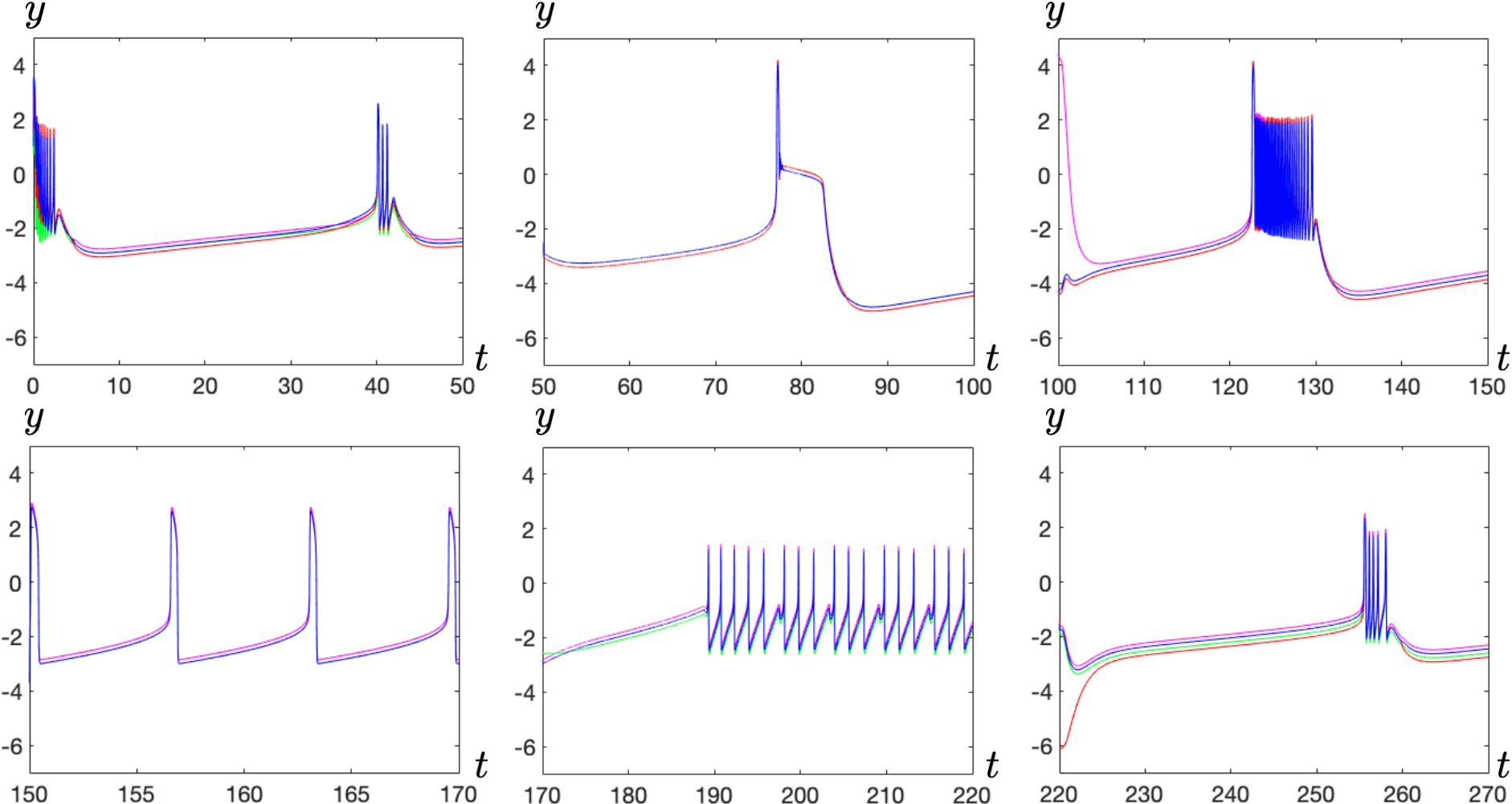}
\caption{Various emergent behaviors depending on the participating agents according to Figure \ref{fig:graph}. Agents 1, 2, 3, and 4 have the colors magenta, green, red, and blue, respectively.}
\label{fig:sim}
\end{center}
\end{figure}

The coupling functions are all chosen initially as $\mu_{ij}(s) = \tan((\pi/2) s)$ and the performance functions are all chosen as $\psi_{ij}(t) = (\pi/2)(0.9\exp(-t)+0.1)$ for $(j,i)\in\cE$.
Then, upon the joining of agent~$1$ at $t=100$, we set $\psi_{14}(t) = (\pi/2)(8.9\exp(-(t-100))+0.1)$.
When agent~$2$ joins at $t=170$, we set $\psi_{24}(t) = (\pi/2)(0.9\exp(-(t-170))+0.1)$.
When agent~$3$ joins at $t=220$, we set $\psi_{32}(t) = (\pi/2)(4.9\exp(-(t-220))+0.1)$. 
These choices ensure that condition~\eqref{eq:lookatme} in Theorem~\ref{thm:funnel} (and Assumptions~\ref{assum:graph} and~\ref{assum:functions} for each connected component) is satisfied at each starting instance by handshake as illustrated in Remark~\ref{Rem:plug-and-play}.
The simulation in Fig.~\ref{fig:sim} is performed in Matlab/Simulink software package with initial conditions $y_i(0) = 1$ and $\zz_i(0) = {\rm col}(0, 0, 0)$, $i \in \mathcal{N}$.

Without coupling, each agent can only converge to an equilibrium, but with coupling they exhibit various emergent behaviors such as spiking pulses and bursting as seen in Fig.~\ref{fig:sim}.
In fact, when all agents are connected, the system is an extension of the FitzHugh-Nagumo model which exhibits the bursting behavior.
Such behavior is utilized in neuromorphic engineering, for instance, to emulate PWM (Pulse Width Modulation)~\cite{sepulchre2022spiking}.
The variety in these periodic behaviors comes from a different limit cycle (hence Assumption~\ref{assum:finite} is satisfied) associated to a different blended dynamics that appears by a different set of agents being connected at each instance (Theorem~\ref{thm:compact}).

\begin{figure}[hbt]
\begin{center}
\includegraphics[width=\columnwidth]{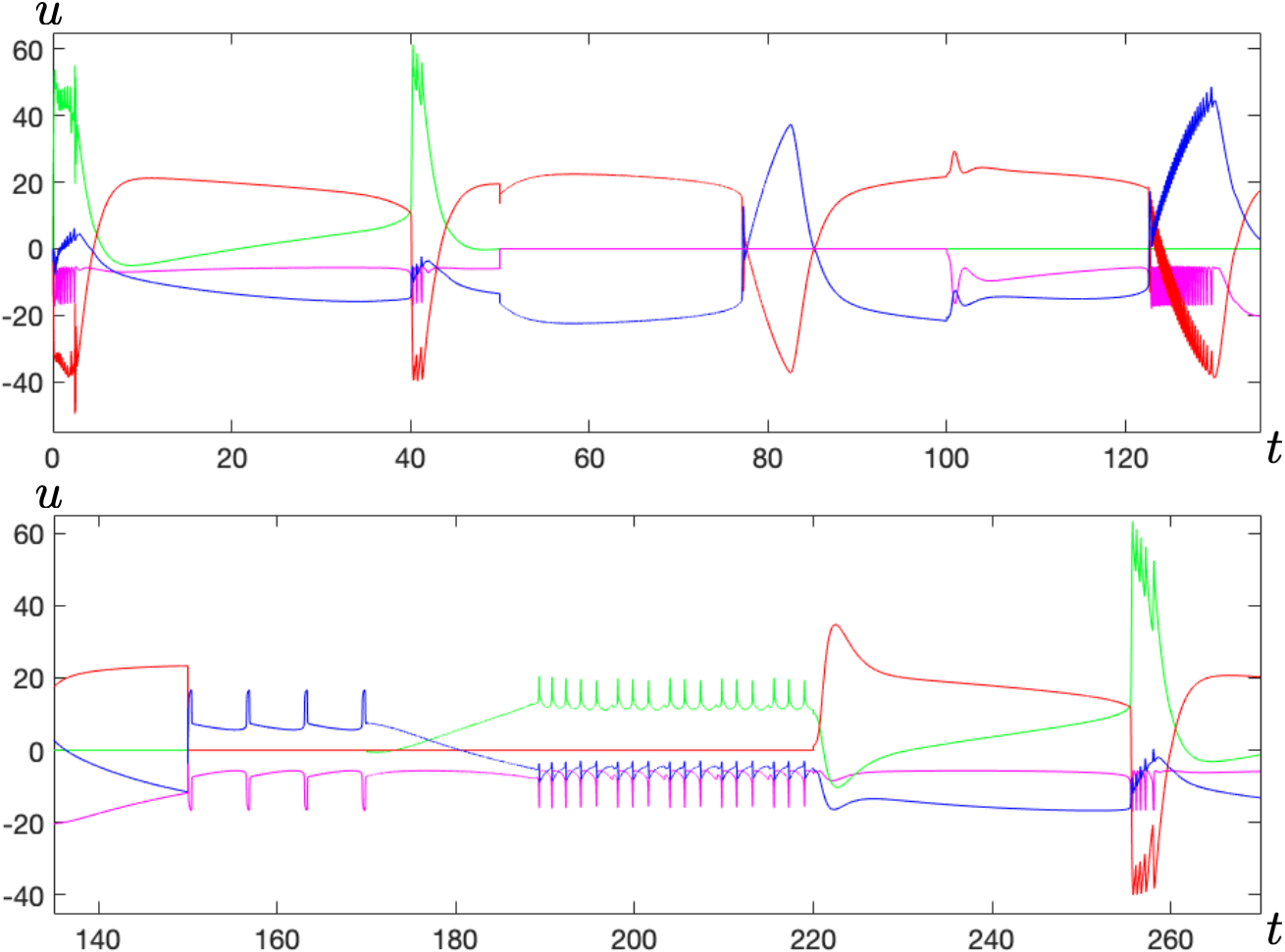}
\caption{$u_1$ (magenta), $u_2$ (green), $u_3$ (red), and $u_4$ (blue).}
\label{fig:funnel}
\end{center}
\end{figure}

Fig.~\ref{fig:funnel} shows that the corresponding inputs are bounded, which in turn implies that the fractions $|\nu_{ij}(t)|/\psi_{ij}(t)$ are uniformly smaller than $1$ and all output differences corresponding to an edge evolve inside the respective funnel (Theorem~\ref{thm:funnel}).

\section{Conclusion}\label{sec:con}

In this paper, we introduced the edge-wise funnel coupling law, which retains all the benign properties of the node-wise funnel coupling law in~\cite{lee2019synchronization}, and exhibits a more straightforward design of the emergent behavior, which is given by the blended dynamics.
The new coupling law is also better suitable for plug-and-play operation.
Future research will focus on the extension of the results to systems with arbitrary relative degree.

\bibliographystyle{ieeetr}
\bibliography{Reference}

\begin{thebibliography}{10}

\bibitem{kim2016robustness}
J.~Kim, J.~Yang, H.~Shim, J.-S. Kim, and J.~H. Seo, ``Robustness of
  synchronization of heterogeneous agents by strong coupling and a large number
  of agents,'' {\em IEEE Trans. Autom. Control}, vol.~61, no.~10,
  pp.~3096--3102, 2016.

\bibitem{panteley2017synchronization}
E.~Panteley and A.~Lor{\'\i}a, ``Synchronization and dynamic consensus of
  heterogeneous networked systems,'' {\em IEEE Trans. Autom. Control}, vol.~62,
  no.~8, pp.~3758--3773, 2017.

\bibitem{jglee18automatica}
J.~G. Lee and H.~Shim, ``A tool for analysis and synthesis of heterogeneous
  multi-agent systems under rank-deficient coupling,'' {\em Automatica},
  vol.~117, p.~108952, 2020.

\bibitem{lee2022design}
J.~G. Lee and H.~Shim, ``Design of heterogeneous multi-agent system for
  distributed computation,'' in {\em Trends in nonlinear and adaptive control}
  (Z.~Jiang, C.~Prieur, and A.~Astolfi, eds.), vol.~488 of {\em Lecture notes
  in control and information sciences}, pp.~83--108, Cham: Springer, 2022.

\bibitem{yun2019initialization}
H.~Yun, H.~Shim, and H.-S. Ahn, ``Initialization-free privacy-guaranteed
  distributed algorithm for economic dispatch problem,'' {\em Automatica},
  vol.~102, pp.~86--93, 2019.

\bibitem{lee2018distributed}
D.~Lee, S.~Lee, T.~Kim, and H.~Shim, ``Distributed algorithm for the network
  size estimation: Blended dynamics approach,'' in {\em 2018 IEEE Conf. Decis.
  Control (CDC)}, pp.~4577--4582, IEEE, 2018.

\bibitem{kim2019completely}
T.~Kim, C.~Lee, and H.~Shim, ``Completely decentralized design of distributed
  observer for linear systems,'' {\em IEEE Trans. Autom. Control}, vol.~65,
  no.~11, pp.~4664--4678, 2019.

\bibitem{shim2015preliminary}
H.~Shim and S.~Trenn, ``A preliminary result on synchronization of
  heterogeneous agents via funnel control,'' in {\em 2015 54th IEEE Conf.
  Decis. Control (CDC)}, pp.~2229--2234, IEEE, 2015.

\bibitem{lee2019synchronization}
J.~G. Lee, S.~Trenn, and H.~Shim, ``Synchronization with prescribed transient
  behavior: Heterogeneous multi-agent systems under funnel coupling,'' {\em
  Automatica}, vol.~141, p.~110276, 2022.

\bibitem{trenn2017edge}
S.~Trenn, ``Edge-wise funnel synchronization,'' {\em Proc. Appl. Math. Mech.},
  vol.~17, no.~1, pp.~821--822, 2017.

\bibitem{leeutility}
J.~G. Lee, T.~Berger, S.~Trenn, and H.~Shim, ``Utility of edge-wise funnel
  coupling for asymptotically solving distributed consensus optimization,'' in
  {\em Proc. Europ. Control Conf.}, (Saint Petersburg, Russia), pp.~911--916,
  2020.

\bibitem{ilchmann2002tracking}
A.~Ilchmann, E.~P. Ryan, and C.~J. Sangwin, ``Tracking with prescribed
  transient behaviour,'' {\em ESAIM: Control Optim. Calc. Var.}, vol.~7,
  pp.~471--493, 2002.

\bibitem{BergLe18}
T.~Berger, H.~H. L{\^e}, and T.~Reis, ``Funnel control for nonlinear systems
  with known strict relative degree,'' {\em Automatica}, vol.~87, pp.~345--357,
  2018.

\bibitem{BergIlch21}
T.~Berger, A.~Ilchmann, and E.~P. Ryan, ``Funnel control of nonlinear
  systems,'' {\em Math. Control Signals Syst.}, vol.~33, pp.~151--194, 2021.

\bibitem{bechlioulis2014robust}
C.~P. Bechlioulis and K.~J. Kyriakopoulos, ``Robust model-free formation
  control with prescribed performance and connectivity maintenance for
  nonlinear multi-agent systems,'' in {\em Proceedings of 53rd IEEE Conf.
  Decis. Control}, pp.~4509--4514, 2014.

\bibitem{bechlioulis2015robust}
C.~P. Bechlioulis and K.~J. Kyriakopoulos, ``Robust model-free formation
  control with prescribed performance for nonlinear multi-agent systems,'' in
  {\em Proceedings of IEEE Int. Conf. Robot. Automation}, pp.~1268--1273, 2015.

\bibitem{bechlioulis2016decentralized}
C.~P. Bechlioulis and G.~A. Rovithakis, ``Decentralized robust synchronization
  of unknown high order nonlinear multi-agent systems with prescribed transient
  and steady state performance,'' {\em IEEE Trans. Autom. Control}, vol.~62,
  no.~1, pp.~123--134, 2017.

\bibitem{verginis2019robust}
C.~K. Verginis, A.~Nikou, and D.~V. Dimarogonas, ``Robust formation control in
  $\mathbb{SE}$ (3) for tree-graph structures with prescribed transient and
  steady state performance,'' {\em Automatica}, vol.~103, pp.~538--548, 2019.

\bibitem{macellari2016multi}
L.~Macellari, Y.~Karayiannidis, and D.~V. Dimarogonas, ``Multi-agent second
  order average consensus with prescribed transient behavior,'' {\em IEEE
  Trans. Autom. Control}, vol.~62, no.~10, pp.~5282--5288, 2017.

\bibitem{mehdifar2020prescribed}
F.~Mehdifar, C.~P. Bechlioulis, F.~Hashemzadeh, and M.~Baradarannia,
  ``Prescribed performance distance-based formation control of multi-agent
  systems,'' {\em Automatica}, vol.~119, p.~109086, 2020.

\bibitem{stamouli2019robust}
C.~J. Stamouli, C.~P. Bechlioulis, and K.~J. Kyriakopoulos, ``Robust dynamic
  average consensus with prescribed performance,'' in {\em Proceedings of 58th
  IEEE Conf. Decis. Control}, (Nice, France), pp.~5420--5425, 2019.

\bibitem{grip2012output}
H.~F. Grip, T.~Yang, A.~Saberi, and A.~A. Stoorvogel, ``Output synchronization
  for heterogeneous networks of non-introspective agents,'' {\em Automatica},
  vol.~48, no.~10, pp.~2444--2453, 2012.

\bibitem{ByrnIsid91a}
C.~I. Byrnes and A.~Isidori, ``Asymptotic stabilization of minimum phase
  nonlinear systems,'' {\em IEEE Trans. Autom. Control}, vol.~36, no.~10,
  pp.~1122--1137, 1991.

\bibitem{Dies17}
R.~Diestel, {\em Graph theory}, vol.~173 of {\em Graduate Texts in
  Mathematics}.
\newblock Berlin-Heidelberg: Springer, 5th~ed., 2017.

\bibitem{ribar2019neuromodulation}
L.~Ribar and R.~Sepulchre, ``Neuromodulation of neuromorphic circuits,'' {\em
  IEEE Trans. Circuits Syst. I: Regular Papers}, vol.~66, no.~8,
  pp.~3028--3040, 2019.

\bibitem{sepulchre2022spiking}
R.~Sepulchre, ``Spiking control systems,'' {\em Proceedings of the IEEE},
  vol.~110, no.~5, pp.~577--589, 2022.

\bibitem{Walt98}
W.~Walter, {\em Ordinary Differential Equations}.
\newblock New York: Springer, 1998.

\bibitem{sontag1996new}
E.~D. Sontag and Y.~Wang, ``New characterizations of input-to-state
  stability,'' {\em IEEE Trans. Autom. Control}, vol.~41, no.~9,
  pp.~1283--1294, 1996.

\end{thebibliography}

\appendix

\section{Graph theoretical lemmas}\label{app:ess_graph_lem}

For technical reasons, regardless of our assumption being that the underlying graph is undirected and connected (Assumption~\ref{assum:graph}), in this section, we present two graph theoretical lemmas, that are essential for our proof of Theorem~\ref{thm:funnel} outlined in Appendix~\ref{app:proof_diff}.
The lemmas are concerned with directed graphs that have no loops.
In Appendix~\ref{app:proof_diff} we will consider directed subgraphs of the original graph that have this property.

Recall that a tuple $(i_0, i_1, \dots, i_l) \in \cN^{l+1}$ is called a \emph{path} (of length $l$) from $i_0$ to $i_l$, if $i_k \in \cN_{i_{k+1}}$ for all $k = 0, \dots, l-1$.
If $i_1,\dots,i_l$ are distinct, then it is called \emph{elementary}.
A \emph{loop} is an elementary path with $i_0 = i_l$.
A node is \emph{isolated}, if it has no incoming/outgoing edges.
A \emph{source} (\emph{sink}) is a node that has no incoming (outgoing) edge.
An isolated node is regarded as a source.
If a graph has no loop and $\cE \neq \emptyset$, then there exist both a source and a sink.
Note that if $\{(i,j), (j,i)\}\in \cE$, then this ``undirected edge'' constitutes a loop $(i,j,i) $ in $\cG$.

\begin{lem1}\label{lem:loop}
Consider a graph $\cG=(\cN,\cE)$ with non-empty $\cE$.
Then $\cG$ has no loop if, and only if, there exists a vector $\boldsymbol{\chi} \in \R^N$ such that $\chi_j - \chi_i > 0$ for all $(j,i) \in \cE$.
\end{lem1}	 

\begin{pf}
(Sufficiency): 
If there is a loop $(i_0, i_1, \ldots, i_l)$ in $\cG$ where $i_0 = i_l$, then we have
\[
0 = \chi_{i_{0}} - \chi_{i_{l}} = \begin{matrix}\sum_{p=0}^{l-1} \end{matrix}(\chi_{i_{p}} - \chi_{i_{p+1}}) > 0
\]
by the assumption, which is a contradiction.

(Necessity): Since there is no loop, every path in the graph is elementary and has a finite length.
Thus, we can define $\tilde{\cN}_k$ as the set of nodes to which a path of maximal length $k$ from a source leads. 
Obviously, $\tilde \cN_0$ is the set of the sources, and there is a maximal length $K$ for all paths in $\cG$.
Then, $\{\tilde{\cN}_k\}_{k=0}^K$ is a partition of $\cN$. 
Now, for each $k=0,\ldots,K$, let $\chi_i := -k$ for all $i \in \tilde \cN_k$.
Then, for all $(j,i) \in \cE$, if $j \in \tilde \cN_k$ for some $k\in \{0,...,K-1\}$ (note that $k=K$ is not possible), then clearly $i \in \tilde \cN_l$ for some $l \in \{k+1,...,K\}$, thus $\chi_j = -k$ and $\chi_i = -l \le -(k+1)$, thus $\chi_j-\chi_i \ge 1 > 0$.
\QEDB
\end{pf}

Let $\cN_\uparrow$ and $\cN_\downarrow$ be the sets of the sources and the sinks, respectively.
Further, let $\cE_\uparrow := \setdef{(j, i) \in \cE}{ j \in \cN_\uparrow }$ and $\cE_\downarrow := \setdef{(j, i) \in \cE}{ i \in \cN_\downarrow }$, which are the outgoing edges from the sources, and the incoming edges to the sinks, respectively.

\begin{lem1}\label{lem:graph_theory}
Consider a graph $\cG=(\cN,\cE)$ with non-empty $\cE$.
If $\cG$ has no loop, then there exist constants $\xi_{ij}>0$ associated with each edge $(j,i) \in \cE$ such that, for all vectors $\boldsymbol{\sigma} \in \R^N$, we have
\begin{equation}\label{eq:equal}
\sum_{(j, i) \in \cE} \xi_{ij}(\sigma_j - \sigma_i) \equiv \sum_{(j, i) \in \cE_\uparrow} \xi_{ij} \sigma_j - \sum_{(j, i) \in \cE_\downarrow} \xi_{ij}\sigma_i .
\end{equation} 
\end{lem1}

\begin{pf}
The graph theoretic interpretation of~\eqref{eq:equal} is the existence of edge weights $\xi_{ij}$, such that for all nodes which are not sinks or sources the sum of the weights of the incoming edges is equal to the sum of the outgoing edges.
We show that, by choosing appropriate edge weights starting from the sources the proof can be concluded.

In the following, we sequentially pick a node $j \in \cN$ and determine $\xi_{ij}$ for all outgoing edges from node $j$.
To this end, let $d_j$ be the out-degree of node $j \in \cN$ (i.e., the number of all outgoing edges), and let $\cE_k := \{ (j,i) \in \cE \;|\; j \in \tilde \cN_k\}$ be the set of all outgoing edges from the nodes in $\tilde \cN_k$, where $\tilde \cN_k$ is as in the proof of Lemma~\ref{lem:loop}.
It is clear that $\{ \cE_k\}_{k=0}^K$ is a partition of $\cE$.
As the first step, for each $(j,i) \in \cE_0$, assign $\xi_{ij} := 1/d_j$.
Regarding $\xi_{ij}$ as the amount of flow through the edge $(j,i)$, this is interpreted as assigning the equally divided outgoing flow from the source.
By this, the incoming flows for all nodes $j \in \tilde \cN_1$ are determined, and thus, we can assign the outgoing flow $\xi_{ij}$ for all $(j,i) \in \cE_1$ as the amount of incoming flow divided by its out-degree:
\begin{equation}\label{eq:kcl}
\xi_{ij} := \frac{1}{d_j} \sum_{l \in \cN_j} \xi_{jl} > 0.
\end{equation}
In this way, we sequentially assign all the outgoing flow for the nodes in $\tilde \cN_k$, $k=0, \ldots, K$, in the increasing order of $k$.
Recalling that $\{\cE_k\}_{k=0}^K$ is a partition of $\cE$, this procedure determines the flow $\xi_{ij}>0$ for all edges in $\cE$.
Then, by construction,~\eqref{eq:equal} holds.
\QEDB
\end{pf}

\section{Proof of Theorem~\ref{thm:funnel}}\label{app:proof_diff}

The proof technique is similar to that of the node-wise funnel coupling case, given in~\cite{lee2019synchronization}, hence we will keep the proof brief.
In this section, we explain the main differences.
For this purpose, we will cite equations from~\cite{lee2019synchronization} as, for example, (3) in~\cite{lee2019synchronization} as (N3).
The full proof is available in the extended version of the paper on arXiv.

First, we show the existence of a unique (local) solution.
Let $q:=Nm+n_1+\ldots+n_N$ and define the relatively open set
\[
    \Omega := \setdef{\!\!\begin{array}{l} (t, \yy_1, \zz_1,\ldots,\yy_N, \zz_N)\\ \in \R_{\ge 0} \times \R^{q}\end{array}\!\!\!}{\!\!\begin{array}{l} \forall\, (j,i) \in \mathcal{E}\ \forall\, p\in\cM:\\  |\nu_{ij}^p| < \psi_{ij}^p(t)\end{array}\!\!\!}
\]
and $\RR:\Omega\to\R^q, (t, \yy_1, \zz_1,\ldots,\yy_N, \zz_N) \newline \mapsto {\rm col}\big(\RR_1,\ldots, \RR_N\big)$ with
\[
    \RR_i = \begin{bmatrix} \FF_i(t,\yy_i,\zz_i) + \sum_{j\in\cN_i} \uu_{ij}(t, \boldsymbol{\nu}_{ij})\\ \ZZ_i(t,\yy_i,\zz_i)\end{bmatrix},
\]
$i\in\cN$.
Then the system~\eqref{eq:eachdyn},~\eqref{eq:funnel_coup} is equivalent to
\begin{align*}
    \dot \xx(t) &= \RR(t,\xx(t)),\\ \xx(t_0) &= {\rm col}\big(\yy_1(t_0),\zz_1(t_0),\ldots,\yy_N(t_0),\zz_N(t_0)\big).
\end{align*}
By assumption we have $\xx(t_0)\in \Omega$ and~$\RR$ is measurable and locally integrable in~$t$ and locally Lipschitz continuous in~$\xx$.
Therefore, by the theory of ordinary differential equations (see e.g.~\cite[\S\,10, Thm.~XX]{Walt98}) there exists a unique maximal solution $\xx:[t_0,\omega)\to\R^q$, $\omega\in(0,\infty]$, of~\eqref{eq:eachdyn} and~\eqref{eq:funnel_coup} which satisfies $(t, \xx(t)) \in \Omega$ for all $t \in [t_0, \omega)$.
Furthermore, the closure of the graph of this solution is not a compact subset of $\Omega$.

Assume that $\omega<\infty$.
Then, different from~\cite{lee2019synchronization}, we find that
\begin{align*}
\cE_+^p(\{\tau_k\}) &:= \setdef{ (j, i) \in \cE }{ \lim_{k \to \infty} \tfrac{\nu_{ij}^p(\tau_k)}{\psi_{ij}^p(\tau_k)} = 1} \\
&\text{ is non-empty for some } p \in \cM, \\
\text{or}& \\
\cE_-^p(\{\tau_k\}) &:= \setdef{ (j, i) \in \cE }{ \lim_{k \to \infty} \tfrac{\nu_{ij}^p(\tau_k)}{\psi_{ij}^p(\tau_k)} = -1} \\
&\text{ is non-empty for some } p \in \cM,
\end{align*}
instead of that $\cI_+(\{\tau_k\})$ is non-empty or $\cI_-(\{\tau_k\})$ is non-empty.
Assuming that $\cE_+^p(\{\tau_k\})$ is non-empty, we will instead show that a contradiction occurs, if the graph $(\cN, \cE_+^p(\{\tau_k\}))$ has a loop.
If $(\cN, \cE_+^p(\{\tau_k\}))$ has no loop, then we will show that it is possible to construct another time sequence $\{\bar{\tau}_k\}$ (based on $\{\tau_k\}$), such that $|\cE_+^p(\{\tau_k\})| < |\cE_+^p(\{\bar{\tau}_k\})|$, similar to (N9).
By repeating the argument, we arrive at a graph $(\cN, \cE_+^p(\{\hat \tau_k\}))$, for some time sequence $\{\hat{\tau}_k\}$, that has a loop, which yields a contradiction as we will show.
Therefore, we conclude that $\omega = \infty$ and~\eqref{eq:res_fun_edge} is achieved.

For convenience, we write $\cE^p$ instead of $\cE_+^p(\{\tau_k\})$ in the following.
Note first that, by the definition of $\cE^p$, there exists $k^*\in\N$ such that for all $k\ge k^*$ and all $(j, i) \in \cE^p$ we have $y_j^p(\tau_k) - y_i^p(\tau_k) = \nu_{ij}^p(\tau_k) > 0$, because $\psi_{ij}^p(t) > 0$ for all $t \in [t_0, \omega)$.
Hence, by Lemma~\ref{lem:loop} the graph $(\cN, \cE^p)$ cannot have a loop.

The remainder of the proof follows as in~\cite{lee2019synchronization}, where we instead use the absolutely continuous function
$$W(t) := \sum_{(j, i) \in \cE^p} \xi_{ij} \nu_{ij}^p(t) = \sum_{(j, i) \in \cE^p} \xi_{ij}\cdot (y_j^p(t) - y_i^p(t))$$
where $\xi_{ij}$ is given by Lemma~\ref{lem:graph_theory} in terms of the graph $(\cN, \cE^p)$.
The sequences $\{\eps_q\}_{q \in \mathbb{N}}$, $\{\tau_{k_q}\}_{q \in \mathbb{N}}$ and $\{s_q\}_{q \in \mathbb{N}}$ are similarly defined as in~\cite{lee2019synchronization}, and similar to (N14) we may conclude that, for some $\bar \xi > 0$,
\begin{equation}\label{eq:Wsq-est}
\forall\, q\in\N:\     \dot{W}(s_q) \ge - \bar \xi \theta_\psi := -\bar{\xi} \sup_{t \ge t_0, (j, i) \in \mathcal{E}, p \in \mathcal{M}} \left|\dot{\psi}_{ij}^p(t)\right|.
\end{equation}

The main difference appears when we arrive at the derivation of (N16).
We have to instead invoke Lemma~\ref{lem:finite} together with Lemma~\ref{lem:graph_theory} in Appendix~\ref{app:ess_graph_lem} for the graph $(\cN,\cE^p)$ to obtain, for almost all $t\in[t_0,\omega)$,
\begin{align}\label{eq:A_up_der_W}
\begin{split}
\dot{W}(t) &\le M_0 + \sum_{(j, i) \in \cE_\uparrow^p} \xi_{ij}\sum_{(l, j) \in \cE} \mu_{jl}^p(t) \\
&\quad - \sum_{(j, i) \in \cE_\downarrow^p} \xi_{ij} \sum_{(l, i) \in \cE} \mu_{il}^p(t),
\end{split}
\end{align}
where $\mu_{kl}^p(t) = \mu_{kl}^p(\nu_{kl}^p(t)/\psi_{kl}^p(t))$ for $(l, k) \in \cE$.
Define the edge sets
\begin{align*}
&\cE_\text{large}^p := \setdef{\!(l, i)\in\cE^p\!}{\exists\, j\in\cN:\ (j, i) \in \cE_\downarrow^p\!} = \cE_\downarrow^p,\\
&\cE_\text{small}^p := \setdef{\!(l, j)\in\cE\!}{\exists\, i\in\cN:\ (j, i) \in \cE_\uparrow^p\!}\\ &\ \ \cup 
 \setdef{(i,l)\in \cE\!}{(l, i) \in \cE \setminus \cE^p,\ \exists\, j\in\cN:\ (j, i) \in \cE_\downarrow^p \!}.
\end{align*}
By definition of $\cE_\uparrow^p$ and $\cE_\downarrow^p$ in Appendix~\ref{app:ess_graph_lem}, we have $\emptyset \neq \cE_\downarrow^p = \cE_\text{large}^p \subset \cE^p$ and $\emptyset \neq \{(i, j)|(j, i) \in \cE_\uparrow^p\} \subset\cE_\text{small}^p \subset \cE \setminus \cE^p$.
The latter holds because from $(j, i) \in \cE_\downarrow^p$, node $i$ is a sink of the graph $(\cN, \cE^p)$, hence $(i, l) \notin \cE^p$ for all $l\in\cN$.
Similarly, if $(j, i) \in \cE_\uparrow^p$, then $j$ is a source of the graph $(\cN, \cE^p)$, hence $(l, j) \notin \cE^p$ for all $l\in\cN$.

Now, since $-\mu_{il}^p(t) = \mu_{li}^p(t)$ for any $(l, i) \in \cE$ by Assumption~\ref{assum:functions}, we can rewrite~\eqref{eq:A_up_der_W} as
\begin{align}\label{eq:A_re_up_der_W}
\dot{W}(t) \le M_0 + \sum_{(l, j) \in \cE_\text{small}^p} \zeta_{jl} \mu_{jl}^p(t) - \sum_{(l, i) \in \cE_\text{large}^p} \zeta_{il}\mu_{il}^p(t)
\end{align}
with positive constants
\[
    \zeta_{jl} = \begin{cases} \sum_{(j,k)\in \cE_\uparrow^p} \xi_{kj} + \sum_{(k,l)\in \cE_\downarrow^p} \xi_{lk}, & (l,j)\in \cE_\text{small}^p,\\
    \sum_{(k,j)\in \cE_\downarrow^p} \xi_{jk}, & (l,j)\in \cE_\text{large}^p.
    \end{cases}
\]
Then, from~\eqref{eq:Wsq-est} and~\eqref{eq:A_re_up_der_W}, we may similarly conclude that
$\sum_{(l,j)\in \cE_\text{small}^p} \max\{\mu_{jl}^p(s_q),0\} \to \infty$ as $q\to\infty$.
Therefore, invoking that $\cE \setminus \cE^p$ is finite, there exist a subsequence $\{\bar{\tau}_k\} = \{s_{q_k}\}$ and an edge $(j^*, i^*) \in \cE_\text{small}^p \subset \cE \setminus \cE^p$ such that $\nu_{i^*j^*}^p(\bar{\tau}_k)/\psi_{i^*j^*}^p(\bar{\tau}_k) \to 1$ as $k \to \infty$.
Consequently, $ \cE_+^p(\{\tau_k\})  \subseteq \cE_+^p(\{s_q\}) \subseteq \cE_+^p(\{\bar{\tau}_k\})$.
Since $(j^*, i^*) \in \cE_+^p(\{\bar{\tau}_k\}) \setminus \cE_+^p(\{\tau_k\})$, the proof concludes.

\section{Proof of Lemma \ref{lem:2}}\label{app:SecSpe}

Construct the function $\gamma$ as follows:
\begin{enumerate}[(i)]
\item set $M_{\xx} := \beta(M_{\xx_0},0) + \hat \gamma(M_{\uu}) > 0$,
\item define $\beta_{M_\xx}(\cdot) := \beta(M_\xx,\cdot)$, which is a strictly decreasing function from $[0,\infty)$ onto $(0,\beta_{M_\xx}(0)]$,
\item for $\eps \in (0,2\beta_{M_\xx}(0)]$ define $\tilde \gamma$ via its inverse
\begin{equation}
\tilde \gamma^{-1}(\eps) := w\left(\beta_{M_\xx}^{-1}\left(\tfrac{\eps}{2}\right)\right) \hat\gamma^{-1}\left(\tfrac{\eps}{2}\right)
\end{equation}
which is a strictly increasing function from $(0,2\beta_{M_\xx}(0)]$ onto $(0,w(0)\hat\gamma^{-1}(\beta_{M_\xx}(0))]$ such that $\lim_{\eps\to0}\tilde \gamma^{-1}(\eps) = 0$.
\item choose any $\gamma \in \Kinf$ such that $\gamma(s) = \tilde \gamma(s)$ for $s \in (0,w(0)\hat\gamma^{-1}(\beta_{M_\xx}(0))]$.
\end{enumerate}
The function $\gamma$ defined above has the following property:
\begin{equation}
\gamma(s) = 2\hat\gamma\left( \tfrac{s}{w(T^*(s))} \right)
\end{equation}
for all $s \in(0,w(0)\hat\gamma^{-1}(\beta_{M_\xx}(0))]$, where $T^*(s) := \beta_{M_\xx}^{-1}(\gamma(s)/2)$ is a strictly decreasing function from $(0,w(0)\hat\gamma^{-1}(\beta_{M_\xx}(0))]$ onto $[0,\infty)$.
(This property is easily proved with $s = \gamma^{-1}(\eps)$.)\\
We now prove the statement of Lemma \ref{lem:2}.
Pick $t > t_0$. 
Then, either (i) $t_0 + T^*(\delta_t) \le t$ or (ii) $t < t_0 + T^*(\delta_t)$ holds. 
If~(i), then for all $s$ such that $t_0 \le t-T^*(\delta_t) \le s \le t$, we have $w(T^*(\delta_t)) \le w(t-s)$.
Therefore, 
\begin{align*}
&\hat{\gamma}\left(\begin{matrix}\sup_{s \in [t - T^*(\delta_t), t)} \end{matrix}  \| \uu(s)\|_\infty\right) \\
&\quad\le \hat{\gamma}\left(\begin{matrix}\sup_{s \in [t - T^*(\delta_t), t)} \end{matrix} \|\uu(s)\|_\infty \tfrac{w(t-s)}{w(T^*(\delta_t))}\right) \\
&\quad\le \hat{\gamma}\left(\tfrac{1}{w(T^*(\delta_t))}\sup_{s \in [t_0, t)}  \|\uu(s)\|_\infty w(t-s)\right) \\
&\quad= \hat{\gamma}\left(\tfrac{\delta_t}{w(T^*(\delta_t))}\right) =  \tfrac{1}{2}\gamma(\delta_t).
\end{align*}
Then, \eqref{eq:sclaim} follows since
\begin{align*}
\|\xx(t)\|_A &\le \beta\left(\|\xx(t - T^*(\delta_t))\|_A, t - (t - T^*(\delta_t))\right) \\
&\quad + \hat{\gamma}\left(\begin{matrix}\sup_{s \in [t - T^*(\delta_t), t)}\end{matrix}  \| \uu(s)\|_\infty\right) \\
&\le \beta\left(M_\xx, T^*(\delta_t)\right) + \tfrac{1}{2}\gamma(\delta_t) = \gamma(\delta_t).
\end{align*}
If (ii), then for all $s$ such that $t_0 \le s \le t$, we have $w(T^*(\delta_t)) \le w(t-s)$.
Thus, we obtain
\begin{align*}
\|\xx(t)\|_A &\le \beta\left(\| \xx(t_0)\|_A, t - t_0\right) + \hat{\gamma}\left(\begin{matrix}\sup_{s \in [t_0, t)}\end{matrix} \| \uu(s)\|_\infty\right) \\
&\le \beta\left(\| \xx(t_0)\|_A, t - t_0\right) \\
&\quad+ \hat{\gamma}\left(\sup_{s \in [t_0, t)} \| \uu(s)\|_\infty \tfrac{w(t-s)}{w(T^*(\delta_t))}\right) \\
&= \beta\left(\| \xx(t_0)\|_A, t - t_0\right) + \hat{\gamma}\left(\tfrac{\delta_t}{w(T^*(\delta_t))}\right)\\
&= \beta\left(\| \xx(t_0)\|_A, t - t_0\right) + \tfrac{1}{2}\gamma\left(\delta_t\right)
\end{align*}
which warrants \eqref{eq:sclaim}.

\newpage

\section{Alternative condition on the reduced blended dynamics (only contained in arXiv-version)}\label{app:SecSpe2}

\begin{lem1}\label{lem:small_gain}
If the system~\eqref{eq:red_blen_inp} with input $(\ee_0,\dots,\dd_N)$ is $\delta$-ISS with $(\beta,\hat{\gamma})$ and if additionally, $\alpha(s) := s - \hat{\gamma}(2s)$ is also a class-$\mathcal{K}_\infty$ function, then the system~\eqref{eq:blended_input} with input $(\ee_1,\dots,\ee_N)$ is $\delta$-ISS. 
\end{lem1}

\begin{pf}
Consider any two solutions of~\eqref{eq:blended_input} denoted as $(\hat{\sss}, \hat{\zz}_1, \dots, \hat{\zz}_N)$ and $(\sss, \zz_1, \dots, \zz_N)$ with inputs $(\hat{\ee}_1, \dots, \hat{\ee}_N)$ and $(\ee_1, \dots, \ee_N)$, respectively.
Fix $i \in \mathcal{N}$.
Then $(\hat\sss,\hat\zz_i)$ is a solution of~\eqref{eq:red_blen_inp} with input ${\rm col}(\hat\ee_i,\hat \ee_1, \dots, \hat\ee_N, \hat\zz_1-\hat\zz_i, \dots, \hat\zz_N-\hat\zz_i)$, and $(\sss,\zz_i)$ is a solution of~\eqref{eq:red_blen_inp} with input ${\rm col}(\ee_i, \ee_1, \dots, \ee_N, \zz_1-\zz_i, \dots, \zz_N-\zz_i)$.
Since~\eqref{eq:red_blen_inp} is $\delta$-ISS with $(\beta, \hat{\gamma})$, we get
\begin{align}\label{eq:dISS17}
\begin{split}
\left\|\!\begin{bmatrix} \hat{\sss}(t) \! - \! \sss(t)\\ \hat{\zz}_i(t) \! - \! \zz_i(t) \end{bmatrix}\! \right\|_\infty \!\!\!\! &\le \beta\!\left(\left\|\!\begin{bmatrix} \hat{\sss}(t_0) \! - \! \sss(t_0)\\ \hat{\zz}_i(t_0) \! - \! \zz_i(t_0) \end{bmatrix}\!  \right\|_\infty \!\!\!\!, t - t_0\!\right) \\
&\quad + \hat{\gamma}\left(\sup_{s \in [t_0, t)} \left\|\!\begin{bmatrix} \tilde{e}(s) \\ 2 \tilde{z}(s)\end{bmatrix}\!\right\|_\infty \right)
\end{split}
\end{align}
where $\tilde{e}(t) := \max_{j \in \mathcal{N}} \|\hat{\ee}_j(t) - \ee_j(t)\|_\infty$ and $\tilde{z}(t) := \max_{j \in \mathcal{N}} \|\hat{\zz}_j(t) - \zz_j(t)\|_\infty$.
Thus, we can conclude that
\begin{align*}
\sup_{s \in [t_0, t)} \left\|\begin{bmatrix} \tilde{s}(s) \\ \tilde{z}(s)\end{bmatrix}\right\|_\infty \!\!\!&\le \beta\left(\left\|\begin{bmatrix} \tilde{s}(t_0) \\ \tilde{z}(t_0)\end{bmatrix}\right\|_\infty \!\!\!, 0\right) \\
&\,\,\,+ \hat{\gamma}\left(2 \!\! \sup_{s \in [t_0, t)} \!\! \tilde{z}(s)\right) + \hat{\gamma}\left(\sup_{s \in [t_0, t)} \!\! \tilde{e}(s)\right)
\end{align*}
where $\tilde{s}(t) := \|\hat{\sss}(t) - \sss(t)\|_\infty$. 
Now, let 
\begin{align*}
\zeta &:= \max\left\{\tilde{s}(t_0), \tilde{z}(t_0), \sup_{s \in [t_0, t)}\tilde{e}(s)\right\}, \\
\bar{\gamma}(s) &:= \alpha^{-1}(2\max\{\beta(s, 0), \hat{\gamma}(s)\}), \quad\quad\quad \forall s \ge 0,
\end{align*}
then we have
\begin{align*}
\sup_{s \in [t_0, t)} \!\! \tilde{z}(s) &\le 2\max\{\beta(\zeta, 0), \hat{\gamma}(\zeta)\} + \hat{\gamma}\left(2 \!\! \sup_{s \in [t_0, t)} \!\! \tilde{z}(s)\right) \\
& = \alpha(\bar{\gamma}(\zeta)) + \sup_{s \in [t_0, t)} \!\! \tilde{z}(s) - \alpha\left(\sup_{s \in [t_0, t)} \!\! \tilde{z}(s)\right),
\end{align*}
and thus, $\sup_{s \in [t_0, t)}  \tilde{z}(s)\le \bar{\gamma}(\zeta)$.
Note that by its definition, $\bar{\gamma} \in \mathcal{K}_\infty$.
We further have
\begin{align}
&\sup_{s \in [t_0, t)}\left\|\begin{bmatrix} \tilde{s}(s) \\ \tilde{z}(s)\end{bmatrix}\right\|_\infty \!\!\! \nonumber \\
&\le \beta\left( 
\left\|\begin{bmatrix} \tilde{s}(t_0) \\ \tilde{z}(t_0)\end{bmatrix}\right\|_\infty
\!\!\!, 0\right)  + \hat{\gamma}\left(\!2\bar{\gamma}\left(\left\| \begin{bmatrix} \tilde{s}(t_0) \\ \tilde{z}(t_0)\end{bmatrix}\right\|_\infty\right)\!\!\right) \nonumber \\
&\quad + \hat{\gamma}\left(\!2 \bar{\gamma}\left(\sup_{s \in [t_0, t)} \!\!\tilde{e}(s)\right)\!\!\right) + \hat{\gamma}\left(\sup_{s \in [t_0, t)} \!\!\tilde{e}(s)\right) \nonumber \\
&\le \bar{\gamma}\left(\left\| \begin{bmatrix} \tilde{s}(t_0) \\ \tilde{z}(t_0)\end{bmatrix}\right\|_\infty\right) +  \bar{\gamma}\left(\sup_{s \in [t_0, t)} \tilde{e}(s)\right) \label{eq:LS8}
\end{align}
where for the last inequality we utilized the fact that $\beta(s, 0) + \hat{\gamma}(2\bar{\gamma}(s))\le \bar{\gamma}(s) $ and $\hat{\gamma}(s)+\hat{\gamma}(2\bar{\gamma}(s)) \le \bar{\gamma}(s)$ for all $s \ge 0$.\footnote{This is because, $\bar{\gamma}(s) - \hat{\gamma}(2\bar{\gamma}(s)) = \alpha(\bar{\gamma}(s)) \ge \beta(s,0),\hat{\gamma}(s)$.}
On the other hand, we can conclude from~\eqref{eq:dISS17} applied to some initial time $T \ge t_0$ (but for the same trajectories), that we have
\begin{align*}
\limsup_{t \to \infty} \left\|\begin{bmatrix} \tilde{s}(t) \\ \tilde{z}(t)\end{bmatrix}\right\|_\infty \!\!\!&\le \hat{\gamma}\left(\sup_{s \in [T, \infty)} \left\|\begin{bmatrix} \tilde{e}(s) \\ 2\tilde{z}(s)\end{bmatrix}\right\|_\infty\right).
\end{align*}
Since the initial time $T$ was arbitrary, we further get
\begin{align*}
\limsup_{t \to \infty} \tilde{z}(t) &\le \!\lim_{T \to \infty} \!\!\hat{\gamma}\left(2\!\!\sup_{s \in [T, \infty)} \!\!\tilde{z}(s)\right) + \hat{\gamma}\left(\sup_{s \in [t_0, \infty)} \!\! \tilde{e}(s)\right) \\
&= \hat{\gamma}\left(2 \limsup_{t \to \infty} \tilde{z}(t)\right) + \hat{\gamma}\left(\sup_{s \in [t_0, \infty)}\!\! \tilde{e}(s)\right),
\end{align*}
hence $\limsup_{t \to \infty} \tilde{z}(t) \le \tilde{\gamma}(\sup_{s \in [t_0, \infty)} \tilde{e}(s))$, where $\tilde{\gamma}(\cdot) := \alpha^{-1}\big( \hat{\gamma}(\cdot)\big)$ is a class-$\mathcal{K}_\infty$ function.
Now, this further implies that
\begin{align}
\limsup_{t \to \infty} \left\|\begin{bmatrix} \tilde{s}(t) \\ \tilde{z}(t)\end{bmatrix}\right\|_\infty &\le \hat{\gamma}\left(2\limsup_{t \to \infty} \tilde{z}(t)\right) + \hat{\gamma}\left(\sup_{s \in [t_0, \infty)} \!\!\tilde{e}(s)\right) \nonumber \\
&\le \tilde{\gamma}\left(\sup_{s \in [t_0, \infty)} \!\! \tilde{e}(s)\right) \label{eq:AG8}
\end{align}
where we utilized the fact that $\tilde{\gamma}(s) = \hat{\gamma}(2\tilde{\gamma}(s)) + \hat{\gamma}(s)$ for all $s \ge 0$.
Then, as in \cite[Theorem 1 ((j) $\Rightarrow$ (a))]{sontag1996new}, $\delta$-ISS of~\eqref{eq:blended_input} follows from~\eqref{eq:LS8} and~\eqref{eq:AG8}.
\QEDB
\end{pf}

\section{Detailed proof of Theorem~\ref{thm:funnel} (only contained in arXiv-version)
}

First, we show the existence of a unique (local) solution.
Let $q:=Nm+n_1+\ldots+n_N$ and define the relatively open set
\[
    \Omega := \setdef{\!\!\begin{array}{l} (t, \yy_1, \zz_1,\ldots,\yy_N, \zz_N)\\ \in \R_{\ge 0} \times \R^{q}\end{array}\!\!\!}{\!\!\begin{array}{l} \forall\, (j,i) \in \mathcal{E}\ \forall\, p\in\cM:\\  |\nu_{ij}^p| < \psi_{ij}^p(t)\end{array}\!\!\!}
\]
and $\RR:\Omega\to\R^q, (t, \yy_1, \zz_1,\ldots,\yy_N,\zz_N) \newline\mapsto {\rm col}\big(\RR_1,\ldots, \RR_N\big)$ with
\[
    \RR_i = \begin{bmatrix} \FF_i(t,\yy_i,\zz_i) + \sum_{j\in\cN_i} \uu_{ij}(t, \boldsymbol{\nu}_{ij})\\ \ZZ_i(t,\yy_i,\zz_i)\end{bmatrix},
\]
$i\in\cN$.
Then the system~\eqref{eq:eachdyn},~\eqref{eq:funnel_coup} is equivalent to
\begin{align*}
    \dot \xx(t) &= \RR(t,\xx(t)),\\ \xx(t_0) &= {\rm col}\big(\yy_1(t_0),\zz_1(t_0),\ldots,\yy_N(t_0),\zz_N(t_0)\big).
\end{align*}
By assumption we have $\xx(t_0)\in \Omega$ and~$\RR$ is measurable and locally integrable in~$t$ and locally Lipschitz continuous in~$\xx$.
Therefore, by the theory of ordinary differential equations (see e.g.~\cite[\S\,10, Thm.~XX]{Walt98}) there exists a unique maximal solution $\xx:[t_0,\omega)\to\R^q$, $\omega\in(0,\infty]$, of~\eqref{eq:eachdyn} and~\eqref{eq:funnel_coup} which satisfies $(t, \xx(t)) \in \Omega$ for all $t \in [t_0, \omega)$.
Furthermore, the closure of the graph of this solution is not a compact subset of $\Omega$.

The proof is done by a contradiction.
Assume that $\omega < \infty$.
This implies that there is a time sequence $\{\tau_k\}$ such that $\tau_k$ is strictly increasing and $\lim_{k \to \infty} \tau_k = \omega$, and 
\begin{align*}
\cE_+^p(\{\tau_k\}) &:= \left\{ (j, i) \in \cE : \lim_{k \to \infty} \frac{\nu_{ij}^p(\tau_k)}{\psi_{ij}^p(\tau_k)} = 1\right\} \\
&\text{ is non-empty for some } p \in \cM, \\
\text{or}& \\
\cE_-^p(\{\tau_k\}) &:= \left\{ (j, i) \in \cE : \lim_{k \to \infty} \frac{\nu_{ij}^p(\tau_k)}{\psi_{ij}^p(\tau_k)} = -1\right\} \\
&\text{ is non-empty for some } p \in \cM.
\end{align*}
Let us first assume that $\cE_+^p(\{\tau_k\})$ is non-empty for some $p \in \cM$.
Then, we will first show that a contradiction occurs if graph $(\cN, \cE_+^p(\{\tau_k\}))$ has a loop.
If graph $(\cN, \cE_+^p(\{\tau_k\}))$ has no loop, we will then show that it is possible to construct another time sequence $\{\bar{\tau}_k\}$ (based on $\{\tau_k\}$), such that  
\begin{align}\label{eq:inc_num_edge}
|\cE_+^p(\{\tau_k\})| < |\cE_+^p(\{\bar{\tau}_k\})|
\end{align}
where the notation $|\cdot|$ implies the cardinality of the set.
By repeating this argument (i.e., by replacing the role of $\{\tau_k\}$ with $\{\bar{\tau}_k\}$), we arrive after finitely many steps at the condition that graph $(\cN, \cE_+^p(\{\tau_k\}))$ has a loop (because $\cE_+^p(\{\tau_k\}) \subset \cE$ and the original graph $(\cN, \cE)$ has a trivial loop represented by an undirected edge), which yields a contradiction.
This means that there is no such sequence $\{\tau_k\}$ that makes $\cE_+^p(\{\tau_k\})$ non-empty for each $p \in \cM$.
Similarly, it can be seen that there is no sequence that makes $\cE_-^p(\{\tau_k\})$ non-empty for each $p \in \cM$.
Therefore, we conclude that there is no such finite time $\omega$, and the control objective~\eqref{eq:res_fun_edge} is achieved.

Let us carry out the above described proof steps.
For convenience, we write $\cE^p$ instead of $\cE_+^p(\{\tau_k\})$ in the following.
Note first that, by the definition of $\cE^p$, there exists $k^*\in\N$ such that for all $k\ge k^*$ and all $(j, i) \in \cE^p$ we have $y_j^p(\tau_k) - y_i^p(\tau_k) = \nu_{ij}^p(\tau_k) > 0$, because $\psi_{ij}^p(t) > 0$ for all $t \in [t_0, \omega)$.
Hence, by Lemma~\ref{lem:loop} the graph $(\cN, \cE^p)$ cannot have a loop.

Now, we continue the proof for the case when the graph $(\cN, \cE^p)$ has no loop, hence becomes a directed subgraph of the original graph $(\cN, \cE)$.
For this purpose, let
$$W(t) := \sum_{(j, i) \in \cE^p} \xi_{ij} \nu_{ij}^p(t) = \sum_{(j, i) \in \cE^p} \xi_{ij}\cdot (y_j^p(t) - y_i^p(t))$$
where $\xi_{ij}$ is given by Lemma~\ref{lem:graph_theory} in terms of the graph $(\cN, \cE^p)$.
Note that $W(t)$ is absolutely continuous, 
\begin{align*}
W(t) &< \sum_{(j, i) \in \cE^p} \xi_{ij} \psi_{ij}^p(t), \quad t \in [t_0, \omega),
\end{align*}
and
\begin{align*}
\lim_{k \to \infty} W(\tau_k) &= \sum_{(j, i)\in \cE^p} \xi_{ij} \psi_{ij}^p(\omega).
\end{align*}
Let us now consider a strictly decreasing sequence $\{\eps_q\} \subset(0,1)$ such that $\lim_{q \to \infty} \eps_q = 0$ and $W(t_0) < (1-\eps_0)\sum_{(j, i)\in \cE^p} \xi_{ij}\psi_{ij}^p(t_0)$.
Choose a subsequence $\{\tau_{k_q}\}_{q \in \mathbb{N}}$ of $\{\tau_k\}$ such that 
\begin{align}\label{eq:cho_subseq}
\forall\, q \in \mathbb{N}:\ W(\tau_{k_q}) \ge \left(1 - \frac{\eps_q}{2}\right)\!\!\sum_{(j, i)\in \cE^p} \!\!\xi_{ij} \psi_{ij}^p(\tau_{k_q}).
\end{align}
Based on this subsequence, we now construct a sequence $\{s_q\}_{q \in \mathbb{N}}$ such that
\begin{align}\label{eq:sel_s_q}
\begin{split}
&s_q := \\
&\max\setdef{ \!\! s \in [t_0, \tau_{k_q}] \!\! }{W(s) = (1 - \eps_q)\!\!\!\!\sum_{(j, i) \in \cE^p} \!\!\xi_{ij}\psi_{ij}^p(s)\!\!}.
\end{split}
\end{align}

By~\eqref{eq:cho_subseq} and~\eqref{eq:sel_s_q}, the sequence $\{s_q\}$ is strictly increasing and $\lim_{q \to \infty} s_q = \omega$.
Moreover, since
$\lim_{q \to \infty} W(s_q) / \sum_{(j, i) \in \cE^p} \xi_{ij}\psi_{ij}^p(s_q) =1$, 
\begin{align}\label{eq:new_seq_conv}
\forall\, (j, i) \in \cE^p:\ \lim_{q \to \infty} \frac{\nu_{ij}^p(s_q)}{\psi_{ij}^p(s_q)} = 1.
\end{align}
In addition, from Assumption~\ref{assum:functions} and from~\eqref{eq:cho_subseq} and~\eqref{eq:sel_s_q}, it follows that $s_q < \tau_{k_q}$ and
\begin{align}\label{eq:low_der_W}
\dot{W}(s_q) \ge (1-\eps_q) \!\! \sum_{(j, i) \in \cE^p} \!\! \xi_{ij}\dot{\psi}_{ij}^p(s_q) \ge - \overline{\xi}\theta_\psi
\end{align}
for all $q\in\N$,\footnote{Without loss of generality, we may choose $\eps_q$ such that $W$ is differentiable at $s_q$, as $W$ is differentiable almost everywhere.} where $\overline{\xi} := \sum_{(j, i) \in \cE^p} \xi_{ij}$ and
$$\theta_\psi := \sup_{t \ge t_0, (j, i) \in \mathcal{E}, p \in \mathcal{M}} \left|\dot{\psi}_{ij}^p(t)\right|.$$

On the other hand, if we compute $\dot{W}$, then we have
\begin{align*}
\dot{W}(t) &= \sum_{(j,i) \in \cE^p} \xi_{ij} ( f_j^p(t, \yy_j, \zz_j) - f_i^p(t, \yy_i, \zz_i)) \\
&\quad + \sum_{(j, i) \in \cE^p} \xi_{ij} \sum_{(l, j) \in \cE} \mu_{jl}^p(t) -  \sum_{(j, i) \in \cE^p} \xi_{ij} \sum_{(l, i) \in \cE} \mu_{il}^p(t)
\end{align*}
for almost all $t\in[t_0,\omega)$, where $\FF_i(t, \yy_i, \zz_i) = {\rm col}(f_i^1(t, \yy_i, \zz_i), \dots, f_i^m(t, \yy_i, \zz_i))$ and $\mu_{kl}^p(t) = \mu_{kl}^p(\nu_{kl}^p(t)/\psi_{kl}^p(t))$, $k \in \cN$, $(l, k) \in \cE$, for simplicity.
We can bound the first sum by $M_0 := \overline{\xi} M_f$, where the constant $M_f$ is such that \begin{align}\label{eq:vec_assum}
\forall\, t \in [t_0, \omega):\ |f_j^p(t, \yy_j(t), \zz_j(t)) - f_i^p(t, \yy_i(t), \zz_i(t))| \le M_f,
\end{align}
whose existence follows from Lemma~\ref{lem:finite} and Assumption~\ref{assum:ind_vec_prop}, because $\omega$ is finite.
Invoking Lemma~\ref{lem:graph_theory} for the edge set $\cE^p$, we therefore have that
\begin{align}\label{eq:up_der_W}
\begin{split}
\dot{W}(t) &\le M_0 + \sum_{(j, i) \in \cE_\uparrow^p} \xi_{ij}\sum_{(l, j) \in \cE} \mu_{jl}^p(t) \\
&\quad - \sum_{(j, i) \in \cE_\downarrow^p} \xi_{ij} \sum_{(l, i) \in \cE} \mu_{il}^p(t)
\end{split}
\end{align}
for almost all $t\in[t_0,\omega)$. 
Let $\cE_\text{large}^p$ be the set of all edges $(l,i)\in\cE$ in~\eqref{eq:up_der_W} such that $(l,i) \in \cE^p$, hence $\lim_{q \to \infty} \mu_{il}^p(s_q) = \infty$;
\begin{align*}
&\cE_\text{large}^p := \setdef{\!(l, i)\in\cE^p\!}{\exists\, j\in\cN:\ (j, i) \in \cE_\downarrow^p\!} = \cE_\downarrow^p.
\end{align*}
By its construction, $\emptyset \neq \cE_\downarrow^p = \cE_\text{large}^p \subset \cE^p$.
Now, since $-\mu_{il}^p(t) = \mu_{li}^p(t)$ for any $(l, i) \in \cE$ by Assumption~\ref{assum:functions}, we can rewrite~\eqref{eq:up_der_W} as
\begin{align}\label{eq:re_up_der_W}
\dot{W}(t) \le M_0 + \sum_{(l, j) \in \cE_\text{small}^p} \zeta_{jl} \mu_{jl}^p(t) - \sum_{(l, i) \in \cE_\text{large}^p} \zeta_{il}\mu_{il}^p(t)
\end{align}
with positive constants
\[
    \zeta_{jl} = \begin{cases} \sum_{(j,k)\in \cE_\uparrow^p} \xi_{kj} + \sum_{(k,l)\in \cE_\downarrow^p} \xi_{lk}, & (l,j)\in \cE_\text{small}^p,\\
    \sum_{(k,j)\in \cE_\downarrow^p} \xi_{jk}, & (l,j)\in \cE_\text{large}^p,
    \end{cases}
\]
where
\begin{align*}
&\cE_\text{small}^p := \setdef{\!(l, j)\in\cE\!}{\exists\, i\in\cN:\ (j, i) \in \cE_\uparrow^p\!}\\ &\ \ \cup 
 \setdef{(i,l)\in \cE\!}{(l, i) \in \cE \setminus \cE^p,\ \exists\, j\in\cN:\ (j, i) \in \cE_\downarrow^p \!}.
\end{align*}
Note that, by the definition of $\cE_\uparrow^p$ and $\cE_\downarrow^p$ in Appendix~\ref{app:ess_graph_lem}, we have $\emptyset \neq \{(i, j)|(j, i) \in \cE_\uparrow^p\} \subset\cE_\text{small}^p \subset \cE \setminus \cE^p$.
This is because from $(j, i) \in \cE_\downarrow^p$, node $i$ is a sink of the graph $(\cN, \cE^p)$, hence $(i, l) \notin \cE^p$ for all $l\in\cN$.
Similarly, if $(j, i) \in \cE_\uparrow^p$, then $j$ is a source of the graph $(\cN, \cE^p)$, hence $(l, j) \notin \cE^p$ for all $l\in\cN$.

Now,~\eqref{eq:low_der_W} and~\eqref{eq:re_up_der_W} yield
$$\sum_{(l, j) \in \cE_\text{small}^p} \!\!\zeta_{jl} \mu_{jl}^p(s_q) \ge \!\!\sum_{(l,i) \in \cE_\text{large}^p}\!\! \zeta_{il} \mu_{il}^p(s_q) - M_0 - \overline{\xi}\theta_\psi =: M_q.$$
Thus, it follows from~\eqref{eq:new_seq_conv} that $M_q \to \infty$ as $q \to \infty$.
Since
$$\sum_{(l, j) \in \cE_\text{small}^p} \zeta_{jl} \mu_{jl}^p(s_q) \le \overline{\zeta} \sum_{(l, j) \in \cE_\text{small}^p}\max\{ \mu_{jl}^p(s_q), 0\}$$
where $\overline{\zeta} := \max_{(l, j) \in \cE_\text{small}^p} \zeta_{jl} > 0$, we have
\begin{align}\label{eq:low_small}
\sum_{(l, j)\in \cE_\text{small}^p} \max\{\mu_{jl}^p(s_q), 0\} \ge \frac{M_q}{\overline{\zeta}}.
\end{align}
Therefore, for each sufficiently large $q$, there is an edge $(j_q, i_q) \in \cE_\text{small}^p \subset \cE \setminus \cE^p$ such that $\mu_{i_qj_q}^p(s_q) \ge M_q /(|\cE_\text{small}^p|\overline{\zeta})$; hence
$$\mu_{i_qj_q}^p\left(\frac{\nu_{i_qj_q}^p(s_q)}{\psi_{i_qj_q}^p(s_q)}\right) \to \infty, \quad \text{i.e.} \quad \frac{\nu_{i_qj_q}^p(s_q)}{\psi_{i_qj_q}^p(s_q)} \to 1.$$
Since $\cE \setminus \cE^p$ is a finite set, there is a subsequence $\{\bar{\tau}_k\} = \{s_{q_k}\}$ such that $(j^*, i^*) = (j_{q_k}, i_{q_k}) \in \cE_\text{small}^p \subset \cE \setminus \cE^p$ and $\nu_{i^*j^*}^p(\bar{\tau}_k)/\psi_{i^*j^*}^p(\bar{\tau}_k) \to 1$ as $k \to \infty$.
Consequently,
$$ \cE_+^p(\{\tau_k\}) \overset{\eqref{eq:new_seq_conv}}{\subseteq} \cE_+^p(\{s_q\}) \subseteq \cE_+^p(\{\bar{\tau}_k\}).$$
By construction, $(j^*, i^*) \in \cE_+^p(\{\bar{\tau}_k\}) \setminus \cE_+^p(\{\tau_k\})$ and we can conclude~\eqref{eq:inc_num_edge} as desired.

\end{document}